\documentclass[prx,amsmath,amssymb,aps,twocolumn]{revtex4-1}

\usepackage{graphicx}
\usepackage{amsmath}
\usepackage{hyperref}
\usepackage[SquareTraceBrackets]{quantum}
\usepackage{bm,natbib,upgreek,amsbsy,mathrsfs,amsfonts}

\DeclareMathOperator{\re}{Re}
\DeclareMathOperator{\im}{Im}
\renewcommand{\identity}{\mathbb{I}}
\DeclareMathOperator{\cov}{Cov}

\begin{document}

\title{Analysis of quantum information processors using quantum metrology}
\author{Mark J. Kandula}
\author{Pieter Kok}
 \email{p.kok@sheffield.ac.uk}
\affiliation{Department of Physics and Astronomy, University of Sheffield, Sheffield S3 7RH, United Kingdom}
\date{\today}

\begin{abstract} 
\noindent
Physical implementations of quantum information processing devices are generally not unique, and we are faced with the problem of choosing the best  implementation. Here, we consider the sensitivity of quantum devices to variations in their different components. To measure this, we adopt a quantum metrological approach, and find that the sensitivity of a device to variations in a component has a particularly simple general form. We use the concept of cost functions to establish a general practical criterion to decide between two different physical implementations of the same quantum device consisting of a variety of components. We give two practical examples of  sensitivities of quantum devices to variations in beam splitter transmitivities: the KLM and Reverse nonlinear sign gates for linear optical quantum computing with photonic qubits, and the enhanced optical Bell detectors by Grice and Ewert \& van Loock. We briefly compare the sensitivity to the diamond distance and find that the latter is less suited for studying the behaviour of components embedded within the larger quantum device.
\end{abstract}

\maketitle

\section{Introduction}
\noindent
Quantum technologies promise dramatic improvements in computation, sensing, and communication, and many efforts are underway to develop it into a mature technology. One of the general challenges is that quantum devices typically need to be extraordinarily precise. We know from quantum fault tolerance theory that models with uncorrelated gate, propagation, and measurement errors may have an error rate of $0.75$\% per element \cite{Raussendorf07}, and it is not known whether more forgiving thresholds exist for equally realistic error models. The tolerances in quantum communication devices are likely less severe, but quantum sensing models are again known to be very susceptible to imperfections in the implementation \cite{Dobrzanski12}. This means that these quantum devices must be fabricated to a very high standard. 

The precision of a device is usually specified in terms of the \emph{fidelity}, which measures how much the actual output state of a device deviates from the intended (ideal) output state. In cases where the ideal output state is pure, the fidelity can be interpreted as the probability of mistaking the actual output state for the ideal output state \cite{Uhlmann76}. It was shown by Myerson \emph{et al.} that a single-shot readout of a qubit in an ion trap can be read out with $99.99$\% fidelity \cite{Lucas08}, and single- and two-qubit gates can achieve fidelities of $99.99$\% and $99.9$\%, respectively \cite{Lucas16}. In other implementations, similar fidelities have been achieved \cite{Ghosh13}. A general method for calculating the fidelity of quantum operations was given by Pedersen \emph{et al.} \cite{Molmer07}, and there is a sizeable literature on measuring deviations from ideal operations using various mathematical techniques \cite{Cui14,Plenio15,Kong17}. However, it was pointed out among others by Sanders, Wallman, and Sanders \cite{Sanders16} that the average gate fidelity is problematic when it comes to assessing the quality of a quantum gate for quantum computing. They show that the gate error rate can be dramatically higher than the fault tolerant threshold even when the average gate fidelity is 99\%. In other words, the average fate fidelity is too optimistic. Even when fault tolerance is not our main concern, such as in various quantum communication protocols, the average gate fidelity may not be the most suitable figure of merit. 

There are often multiple ways to implement a device, sometimes with dramatically different susceptibilities to variations in the device's components \cite{Crickmore16}. Given additional constraints such as costs, it is not clear \emph{a priori} how an array of fidelities associated with different component variations should be combined into a single number that can be used to identify the best way to implement the device. Another technical complication is that the fidelity is a \emph{function} of device parameters, rather than a single number. In order to obtain a meaningful value for the fidelity, we must choose some non-zero deviation of the device parameters since for zero deviations the fidelity will by definition be equal to unity. This choice of deviation introduces a level of arbitrariness into the metric that we wish to avoid. Instead, we want a single number for each component (operating perfectly) that indicates the sensitivity of the device to deviations in that component.

\begin{figure}[t!]
\centering
 \includegraphics[width=8.5cm]{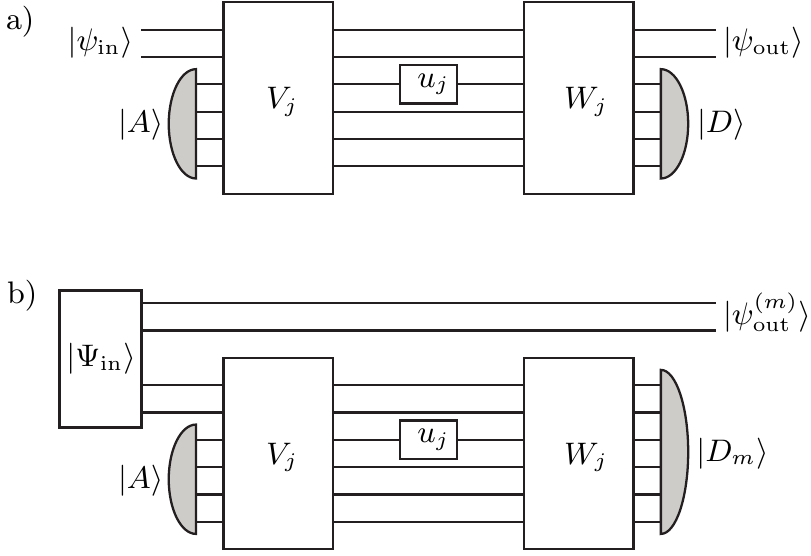}
 \caption{Decomposition of a quantum device into unitary components, specifically highlighting the component $u_j$. The horizontal lines may be qubits or optical modes, depending on the implementation. a) A gate $\ket{\psi_{\rm out}} = \mathscr{U}_g\ket{\psi_{\rm in}}$ with input state $\ket{\psi_{\rm in},A}$ and post-selected on a projection on $\ket{D}$; b) a quantum measurement device that has no output state, but gives a classical output ``$m$'', indicated by the state $\ket{D_m}$. Entangling the input $\ket{\psi_{\rm in}}$ with an auxiliary system prior to the device's operation ($\ket{\Psi_{\rm in}}$) allows us to apply the techniques for a) to measurements.}
 \label{fig:device}
\end{figure}

In this paper, we propose a method for testing the sensitivity of quantum devices that is not based on the fidelity. We use a method from quantum metrology to approach this problem, where the output state of the quantum device carries information about the characteristics of the device's components. This leads to the definition of the \emph{sensitivity} of the device to variation in a component, and for a multi-component device we will obtain a sensitivity matrix. Together with a cost function for the different components, this sensitivity matrix provides a clear metric for the performance of different architectures for the same quantum device.

This paper is organised as follows: in section~\ref{sec:sensitivity} we introduce the {sensitivity} for a device component. To realise this, we divide quantum devices into two categories, namely gates and measurement devices. The latter differs from the former in that there is no output state to the device. In section~\ref{sec:examples} we demonstrate how the sensitivity works for two incarnations of the nonlinear sign gate in linear optical quantum computing \cite{klm01,Crickmore16}, and for two implementations of the enhanced optical Bell measurement \cite{Grice11,Ewert14}. In section~\ref{sec:design} we bring together the sensitivities for different components into a single metric that tests different implementations of a quantum device. In section~\ref{sec:diamond} we briefly comment on the relation between our sensitivity and the diamond norm. We conclude our discussion in section~\ref{sec:conclusions}.

\section{Defining the {Sensitivity} for components of quantum devices}\label{sec:sensitivity}\noindent
We wish to consider the component sensitivity of two kinds of quantum devices. First, we consider quantum gates that have an input and an output, and which may be based on post-selection of auxiliary quantum systems (e.g., qubits or photons). This situation is depicted in Fig.~\ref{fig:device}a. As an example of this type of device, we will consider the nonlinear sign gate of linear optical quantum computing with photonic qubits \cite{Crickmore16,klm01}. Second, we consider complex detection devices that use quantum gates to implement the desired observable. In this situation there is no surviving quantum state that can be used to track variations in components. To remedy this, we use entangled input states that allow us to define the action of a measurement in terms of a surviving quantum state \cite{kok01b}. This situation is depicted in Fig.~\ref{fig:device}b. As an example of this type of device, we will consider enhanced Bell measurements \cite{Grice11,Ewert14}.

\subsection{Quantum gates}\noindent
Consider a quantum gate $g$ described by the unitary evolution $\ket{\psi_{\rm out}} = \mathscr{U}_g\ket{\psi_{\rm in}}$, where the evolution can be post-selected using an auxiliary input state $\ket{A}$ and a detected state $\ket{D}$ (see Fig.~\ref{fig:device}). The detected state may be one of a family of states that herald a successful gate. The intervening evolution can often be decomposed in terms of $N$ smaller unitary operations $U = \prod_{j=1}^N u_j$. These $u_j$ are the physical components of the quantum device generated by a Hamiltonian $H_j$:
\begin{align}
 u_j = \exp(-i\theta_j H_j)\, ,
\end{align} 
with $\theta_j$ the component parameter whose value determines the gate operation. In general, the practical gate operation is more accurately described by a completely positive map that allows for imperfections in the device, but here we are interested in the ideal device and how deviations in the components affect the gate. While a more general discussion is certainly possible, it would also obscure some of the more intuitive aspects of this work. 

After normalisation, the output of the device can be written as 
\begin{align}\label{eq:gate}
 \ket{\psi_{\rm out}} = \frac{1}{\sqrt{p}} \braket{D|U|\psi_{\rm in},A}\, ,
\end{align}
where $p = \norm{\braket{D|U|\psi_{\rm in},A}}^2$ is the probability of success of the quantum device that implements the operation $\mathscr{U}_g$. Suppose we are interested in the $j^{\rm th}$ component of the device, denoted by $u_j$. Define 
\begin{align}
 V_j = \prod_{k=1}^{j-1} u_k \qquad\text{and}\qquad 
 W_j = \prod_{k=j+1}^{N} u_k\, . 
\end{align}
Then we can decompose the output state as (see Fig.~\ref{fig:device}a):
\begin{align}
 \ket{\psi_{\rm out}} = \frac{1}{\sqrt{p}} \braket{D|W_j u_j V_j|\psi_{\rm in},A}\, .
\end{align}
We can treat the sensitivity of the device to variations in $u_j$ as an estimation problem of the parameter $\theta_j$ that characterises the component $u_j$. To this end we use the output state $\ket{\psi_{\rm out}}$ as the basis for the estimation procedure. This state is already post-selected on the correct measurement outcome $\Pi_D \equiv \ket{D}\bra{D}$. This is consistent with the operation of the gate, where the quantum computer trusts that upon getting the measurement outcome ``$D$'' the gate does what it is supposed to do.

Fortunately, we do not explicitly have to perform a complicated estimation procedure. Instead, we can calculate the average amount of information about $\theta_j$ that is contained in the output state $\ket{\psi_{\rm out}}$. If the output state is very sensitive to variations in $\theta_j$ (the aspect we are trying to capture), then it must by definition vary strongly when the value of $\theta_j$ changes. The variation of the output state with $\theta_j$ is quantified by the quantum Fisher information $\smash{I_Q^{(j)}}$, according to \cite{braunstein95}
\begin{align}
 I_Q^{(j)} = \braket{\partial_j{\psi}_{\rm out}|\partial_j{\psi}_{\rm out}} - \Abs{\braket{\psi_{\rm out}|\partial_j{\psi}_{\rm out}}}^2\, ,
\end{align}
where $\partial_j$ is the partial derivative with respect to $\theta_j$. When we define 
\begin{align}
 \ket{\phi_{\rm in}} = u_j V_j \ket{\psi_{\rm in},A}
  ~\text{and}~ 
 \ket{\phi_{\rm out}} = W_j^\dagger \ket{\psi_{\rm out},D} \, ,
\end{align}
the derivative of the output state is compactly written as 
\begin{align}
 \ket{\smash{\partial_j{\psi}_{\rm out}}} = \frac{-i}{\sqrt{p}} \braket{D|W_j H_j|\phi_{\rm in}} +\frac12 \left( \partial_j \log p \right) \ket{\psi_{\rm out}}\, ,
\end{align}
where $H_j$ is the generator of translations in $\theta_j$. We can then explicitly calculate the quantum Fisher information. First we calculate
\begin{align}
 \braket{\partial_j{\psi}_{\rm out}|\partial_j{\psi}_{\rm out}}  = ~ & \frac{1}{p} \Norm{\braket{D|W_j H_j|\phi_{\rm in}}}^2 + \frac14 \left( \partial_j \log p \right)^2  \cr & - \frac{\partial_j \log p}{\sqrt{p}}\, \im\braket{\phi_{\rm in}|H_j|\phi_{\rm out}}\, ,
\end{align}
and 
\begin{align}
 \braket{\psi_{\rm out}|\partial_j{\psi}_{\rm out}} = \frac{-i}{\sqrt{p}} \braket{\phi_{\rm out}|H_j|\phi_{\rm in}} +\frac12 \left( \partial_j \log p \right) \, .
\end{align}
From this, we find that 
\begin{align}
 I_Q^{(j)} = \frac{1}{p} \Norm{\Braket{D|W_j H_j|\phi_{\rm in}}}^2 -  \frac{1}{p} \Abs{\Braket{\phi_{\rm out}|H_j|\phi_{\rm in}}}^2\, .
\end{align}
We can clean up this expression by inserting a resolution of the identity $\identity = \smash{\ket{\psi_{\rm out}}\bra{\psi_{\rm out}} + \sum_k \ket{k}\bra{k}}$ in the first term of $I_Q^{(j)}$, where the orthonormal states $\ket{k}$ complete $\ket{\psi_{\rm out}}$ to form an orthonormal basis of the output Hilbert space. We find that 
\begin{align}
 I_Q^{(j)} = \frac{1}{p} \sum_{k} \Abs{\Braket{\phi_k| H_j|\phi_{\rm in}}}^2\, ,
\end{align}
where we defined $\ket{\phi_k}\equiv W_j^\dagger \ket{k,D}$. We can understand this expression as the quadratic sum over the weak values of the generator $H_j$ given the input state $\ket{\psi_{\rm in},A}$ and the output states $\ket{k,D}$ that are \emph{orthogonal} to the intended output state $\ket{\psi_{\rm out},D}$. The success probability $p$ of the quantum device is a common factor in $\smash{I_Q^{(j)}}$ and does not play a role in the determination of the component sensitivity of a device (although it is important to include this factor when comparing the sensitivity of components in \emph{different} devices with different $p$). In general, $p$ changes when the component $u_j$ changes, and this can in principle be used in an estimation procedure of $\theta_j$. However, we post-select the state on the detection outcome $\Pi_D$, which means we have already discarded the information about the success rate of the quantum device. This is consistent with the normal operation of the gate $\mathscr{U}_g$.

The sensitivity $S_j$ of the quantum device to components $u_j$ can now be defined as 
\begin{align}\label{eq:sensitivity}
 S_j \equiv \sum_{k} \abs{\braket{\phi_k| H_j|\phi_{\rm in}}}^2\, .
\end{align}
While this is an elegant theoretical expression that gives a clear intuitive meaning for $S_j$, for practical purposes it may be beneficial to express $S_j$ instead as
\begin{align}
 S_j = \braket{\phi_{\rm in}|H_j K_D H_j |\phi_{\rm in}} - \abs{\braket{\phi_{\rm out}|H_j|\phi_{\rm in}}}^2\, ,
\end{align}
using $K_D = W_j^\dagger (\identity \otimes \Pi_D) W_j$. This expression does not require the construction of the complementary basis states $\ket{k}$. It also holds for gates that rely on higher rank post-selection described by projectors $\Pi_D$, such as for example the double heralding procedure for creating entangled networks \cite{Barrett05}.

To determine a general, non-state-specific sensitivity of a device, we can average $S_j$ over all possible input states. Alternatively, we can take as a standard input state an equal superposition of the eigenstates of $\mathscr{U}_g$, which is computationally much more straightforward.

\subsection{Quantum measurement devices}\noindent
Next, we consider quantum measurement devices, as shown in Fig~\ref{fig:device}b. The situation is slightly more complicated than the sensitivity for gate components, since there are typically multiple detection outcomes $m$, corresponding to projections onto $\ket{D_m}$ (which in turn are generally projections onto a subspace of the output space). The corresponding surviving quantum state $\ket{\smash{\psi_{\rm out}^{(m)}}}$ is defined by
\begin{align}
 \ket{\smash{\psi_{\rm out}^{(m)}}} = \frac{1}{\sqrt{p_m}} \braket{D_m|U|\Psi_{\rm in},A}\, ,
\end{align}
where the input state $\ket{\Psi_{\rm in}}$ is a maximally entangled state that allows us to relate the measurement outcome to an output state that can be used to define the sensitivity:
\begin{align}
 \ket{\Psi_{\rm in}} = \frac{1}{\sqrt{d}} \sum_k \ket{B_k,B_k}\, ,
\end{align}
with $d$ the dimension of the input state space of the measurement device. The states $\ket{B_k}$ are the orthonormal eigenstates of the observable measured in the measurement device \cite{kok01b}. 

To calculate the sensitivity of the $j^{\rm th}$ component of the measurement device, parameterised by $\theta_j$, we again calculate the quantum Fisher information of $\theta_j$ in the output state $\ket{\smash{\psi_{\rm out}^{(m)}}}$. Clearly, this will be different for different outcomes $m$, and we define the quantum Fisher information $\smash{I_Q^{(j,m)}}$ for each component $j$ and measurement outcome $m$. The total quantum Fisher information for the $j^{\rm th}$ component is then the weighted sum over all measurement outcomes
\begin{align}
 I_Q^{(j)} = \sum_{m\neq m_f} p_m \, I_Q^{(j,m)}\, .
\end{align}
One subtlety that we will encounter in the next section is that sometimes there are outcomes $m_f$ of the measurement device that indicate the measurement has \emph{failed} to produce a useful outcome. There may still be information in $\ket{\smash{\psi_{\rm out}^{(m)}}}$, but since these outcomes (and any post-selection based on these outcomes) are discarded in normal operation of the device, deviations in $\ket{\smash{\psi_{\rm out}^{(m_f)}}}$ have no effect on the device operation and we must \emph{not} include $\smash{I_Q^{(j,m_f)}}$ in the calculation of the sensitivity.

Proceeding with the calculation of $\smash{I_Q^{(j,m)}}$, we use that 
\begin{align}
 I_Q^{(j,m)} = \Braket{\partial_j{\psi}^{(m)}_{\rm out}|\partial_j{\psi}^{(m)}_{\rm out}} - \Abs{\Braket{\psi^{(m)}_{\rm out}|\partial_j{\psi}^{(m)}_{\rm out}}}^2\, .
\end{align}
Following the same method as in the previous section, we find that 
\begin{align}
 I_Q^{(j,m)} = ~& \frac{1}{p_m} \Braket{\phi_{\rm in}| H_j W_j^\dagger \left(\Pi_m \otimes \identity\right) W_j H_j|\phi_{\rm in}} \cr & - \Abs{\Braket{\phi_{\rm out}^{(m)}|H_j|\phi_{\rm in}}}^2\, ,
\end{align}
where $\Pi_m$ is the projector onto the subspace associated with the state $\ket{D_m}$, which may have rank greater than one, and the states $\ket{\smash{\phi_{\rm out}^{(m)}}}$ and $\ket{\phi_{\rm in}}$ are defined as
\begin{align}
 \ket{\phi_{\rm in}}  = u_j V_j \ket{\Psi_{\rm in},A} ~\text{and}~
 \ket{\psi_{\rm out}^{(m)}}  = W_j^\dagger \ket{\psi_{\rm out}^{(m)},D_m} .
\end{align}
The unitary evolutions $u_j$, $V_j$ and $W_j$ are defined as in Fig.~\ref{fig:device}b. We can insert a resolution of the identity into the first term:
\begin{align}
 \identity = \ket{\psi_{\rm out}^{(m)}}\bra{\psi_{\rm out}^{(m)}} + \sum_k \ket{\xi_k^{(m)}}\bra{\xi_k^{(m)}}\, ,
\end{align}
for some orthonormal set $\{ \ket{\smash{\xi_k^{(m)}}} \}$ that span the subspace $\identity - \ket{\smash{\psi_{\rm out}^{(m)}}}\bra{\smash{\psi_{\rm out}^{(m)}}}$, and this leads to 
\begin{align}
 I_Q^{(j,m)} = \frac{1}{p_m} \sum_k \Abs{\Braket{\phi^{(m)}_k | H_j | \phi_{\rm in}}}^2\, ,
\end{align}
with $\ket{\smash{\phi^{(m)}_k}} = W_j^\dagger \ket{\smash{\xi_k^{(m)},D_m}}$. The sensitivity then becomes
\begin{align}
 S_j & = \sum_{k=1}^{d-1} \sum_{m\neq m_f}   \Abs{\Braket{\phi^{(m)}_k | H_j | \phi_{\rm in}}}^2\cr & =   \sum_{k=1}^{d-1} \sum_{m\neq m_f} \braket{\phi_{\rm in} | H_j \left( \widetilde{\Pi}_m \otimes \identity \right) H_j |\phi_{\rm in}}\, ,
\end{align}
where $\widetilde{\Pi}_m \equiv W_j^\dagger \Pi_m W_j$. Two explicit examples of the beam splitter sensitivity of optical Bell measurements are given in section \ref{sec:Bell}.

One may notice that the sensitivity, as measured by the quantum Fisher information, carries units of the inverse-squared of $\theta_j$. When the $\theta_j$ refer to different components in a device, these units may be different, and a straight comparison will not be possible. Indeed, this is a key problem in constructing a single metric for different implementations of a quantum device, and we will return to this issue in section~\ref{sec:design}.

\subsection{Variations in sources and detectors}\noindent
The discussion so far has been restricted to unitary elements in a quantum device, but in practice we will also have to include variations in the auxiliary input states and the detectors. These can be included as unitary evolutions also. For example, an auxiliary input state $\ket{A}$ may be transformed into a different input state $\ket{A(\bm\theta_A)}$ according to a unitary evolution
\begin{align}
 \ket{A(\bm\theta_A)} = u_A(\bm\theta_A) \ket{A}
\end{align}
with
\begin{align}
 u_A(\bm\theta) = \exp\left( -\frac{i}{\hbar}  \bm{H}_A \cdot \bm\theta_A \right)\, ,
\end{align}
where $\bm\theta_A$ is a vector of parameters associated with the evolution from $\ket{A}$ to $\ket{A(\bm\theta_A)}$ generated by $\bm{H}_A$. In the ideal case, $\bm\theta_A=0$ and we recover the original auxiliary state. The sensitivity of the device to the auxiliary input state with respect to a particular variation $\theta_{A,j}$, the $j^{\rm th}$ entry of $\bm\theta_A$, is then defined as 
\begin{align}
 S_{A,j} = \braket{\psi_{\rm in},A| H_{A,j} U^\dagger (\identity \otimes \Pi_D) U H_{A,j} |\psi_{\rm in},A}\, ,
\end{align}
where $H_{A,j}$ is the  $j^{\rm th}$ entry of $\bm{H}_A$, and $U = W_j u_j V_j$ is the unitary evolution defined in Eq.~(\ref{eq:gate}) and Fig.~\ref{fig:device}. Often, a physical imperfection in the source will lead to auxiliary input states that are mixed. However, due to the convex nature of the quantum Fisher information, we need only consider the effect of unitary deviations from the pure ancilla state.

Similarly, for imperfect detectors there is a way of calculating the sensitivity using the technique developed above. In the most general terms, an imperfect detector does not measure the exact observable $M_D$, but instead some rotated observable 
\begin{align}
 M_D(\bm\theta_D) = u_D M_D u_D^\dagger\, ,
\end{align}
where $u_D = \exp(-i \bm{H}_D \cdot\bm\theta_D/\hbar)$ is the unitary evolution that rotates the eigenbasis of $M_D$ to the eigenbasis of $M_D(\bm\theta_D)$. The sensitivity of the device to detector imperfections with respect to a particular variation $\theta_{D,j}$ is then defined as 
\begin{align}
 S_{D,j} = \braket{\psi_{\rm in},A| U^\dagger (\identity \otimes H_{D,j} \Pi_D H_{D,j} )  U |\psi_{\rm in},A}\, ,
\end{align}
where $H_{D,j}$ is the  $j^{\rm th}$ entry of $\bm{H}_D$. This allows for a complete analysis of which variations in the observable are most detrimental to the device operation.

\subsection{Stochastic noise}\noindent
So far, we have considered sensitivities to systematic errors in the device components. A natural question is whether we can include stochastic noise in the analysis of our quantum devices. Stochastic noise models generally describe the situation where some of the parameters $\theta_j$ are fluctuating over time, rather than offset by some amount from the ideal value. It is important to note that the sensitivity defined here is an intrinsic property of each device component, and does not change with the type of deviation from the ideal values of the parameters, systematic or stochastic. Therefore, the device component analysis presented in this section is complete. The different types of errors and noise that a device exhibits is included instead in the metric used to decide between implementations. We will return to this aspect in section~\ref{sec:design}.

\begin{figure}[t!]
\centering
 \includegraphics[width=6cm]{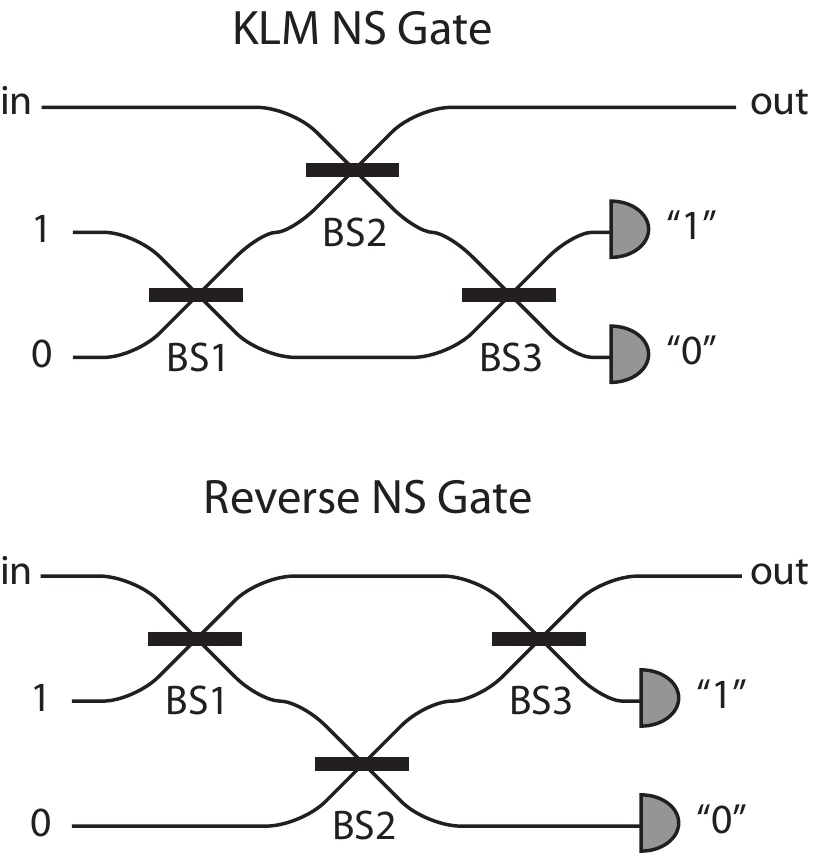}~~
 \caption{The nonlinear sign (NS) circuit for linear optical quantum computing. There are multiple versions of this circuit that are equivalent in terms of the success probability, number of optical elements, auxiliary photons and detectors, but they exhibit inequivalent behaviour in the presence of variations in the components. Here, we consider the KLM NS gate (top) introduced in Ref.~\cite{klm01}, and the Reverse NS gate (bottom) introduced in Re.~\cite{Crickmore16}.}
 \label{fig:ns}
\end{figure}

\section{Examples}\label{sec:examples}\noindent
To demonstrate the sensitivity measure, we consider several examples. We calculate the sensitivities to variations in the beam splitters in nonlinear sign (NS) gates for photonic linear optical quantum computing. These devices fall in the category of quantum gates. Next, we calculate the sensitivities to variations in the beam splitters in two types of optical Bell detectors, which fall in the category of quantum measurement devices. We compare the sensitivities to the device fidelity in order to show that the sensitivity behaves as expected. The dependence of optical gates on their constituent optical components has been studied before \cite{glancy02,ralph02a,lund03,rohde05a,rohde05b,rohde05c,Clements16}, but to our knowledge a unifying model for comparing implementations in arbitrary quantum device architectures has not yet been proposed.

\subsection{The Nonlinear Sign gate in LOQC}\noindent
The NS gate is a key component in the original proposal for linear optical quantum computing with photonic qubits by Knill, Laflamme and Milburn in 2001 \cite{klm01}. It is a probabilistic gate that implements the unitary evolution
\begin{align}
 \alpha\ket{0} + \beta\ket{1}+\gamma\ket{2} \to \alpha\ket{0} + \beta\ket{1} - \gamma\ket{2}\, ,
\end{align}
with $\ket{n}$ denoting single mode optical Fock states. The success probability is one quarter, and there are several inequivalent ways to implement the NS gate, two of which are shown in Fig.~\ref{fig:ns}. The beam splitter values in the two implementations are different, and we reported in Ref.~\cite{Crickmore16} that the gate operation in the presence of variations in the beam splitter reflectivities depends strongly on the implementation.

\begin{figure}[t!]
\centering
 \includegraphics[width=4cm]{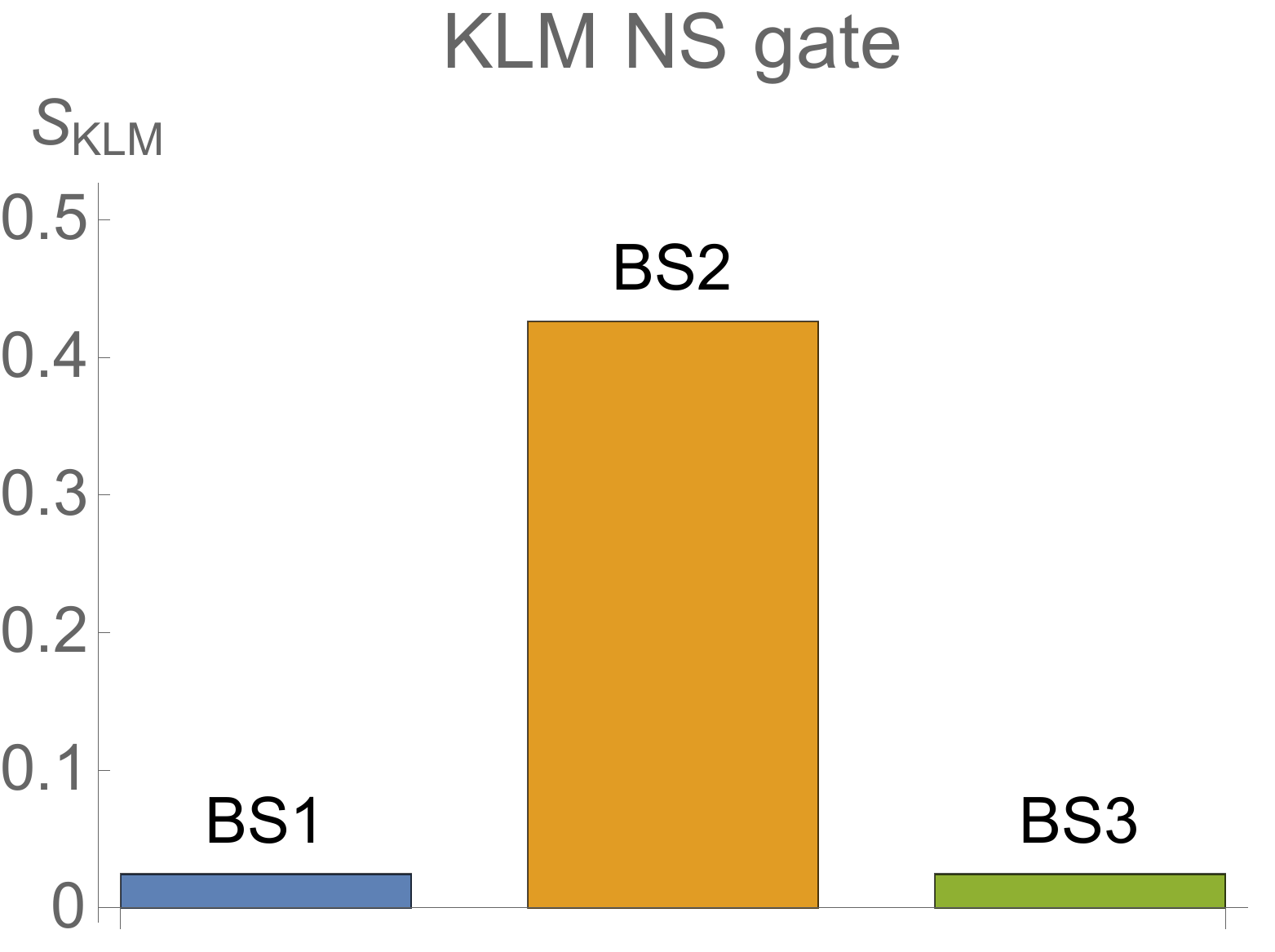} 
 \includegraphics[width=4.5cm]{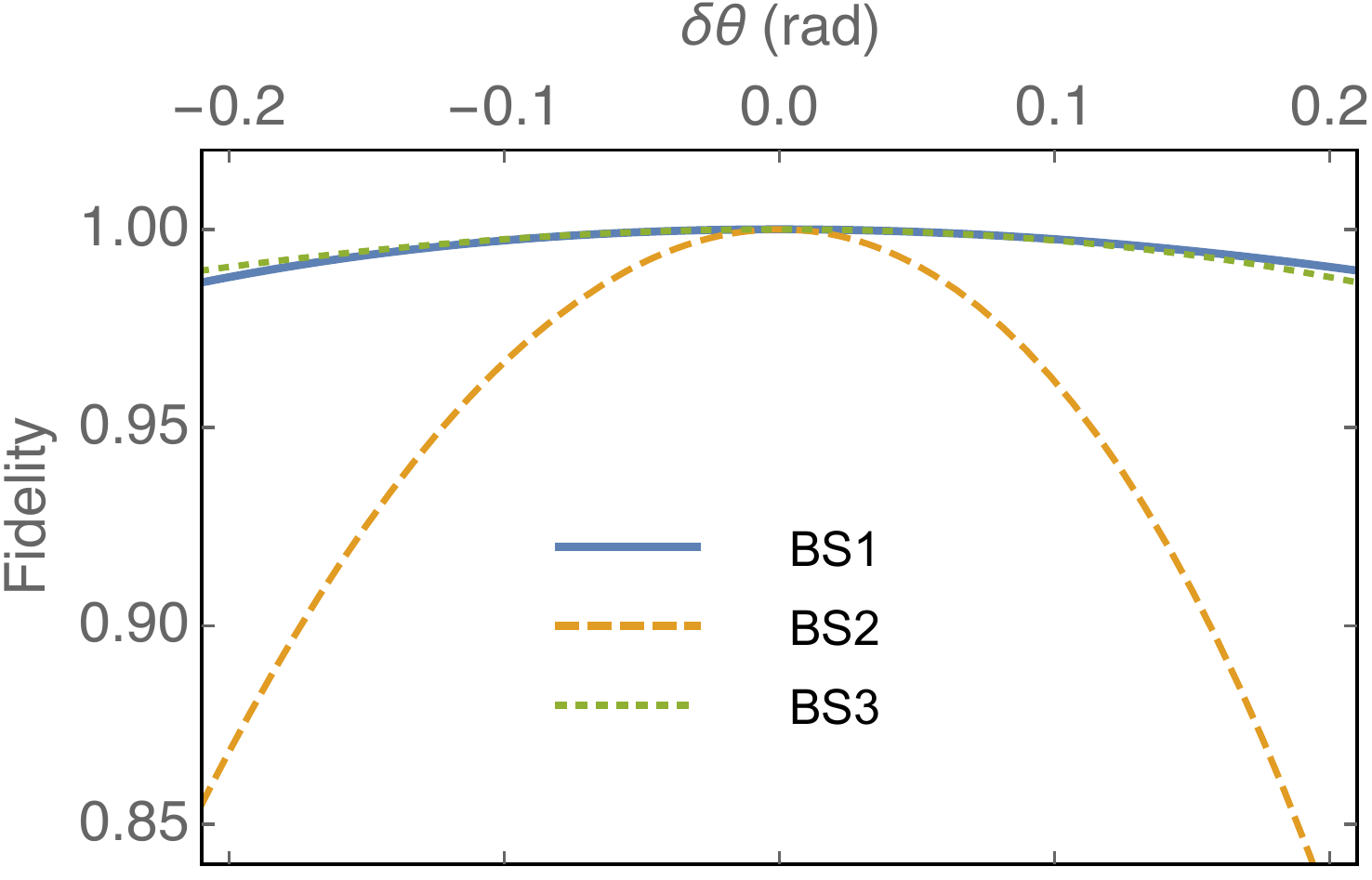} \\ \bigskip
 \includegraphics[width=4cm]{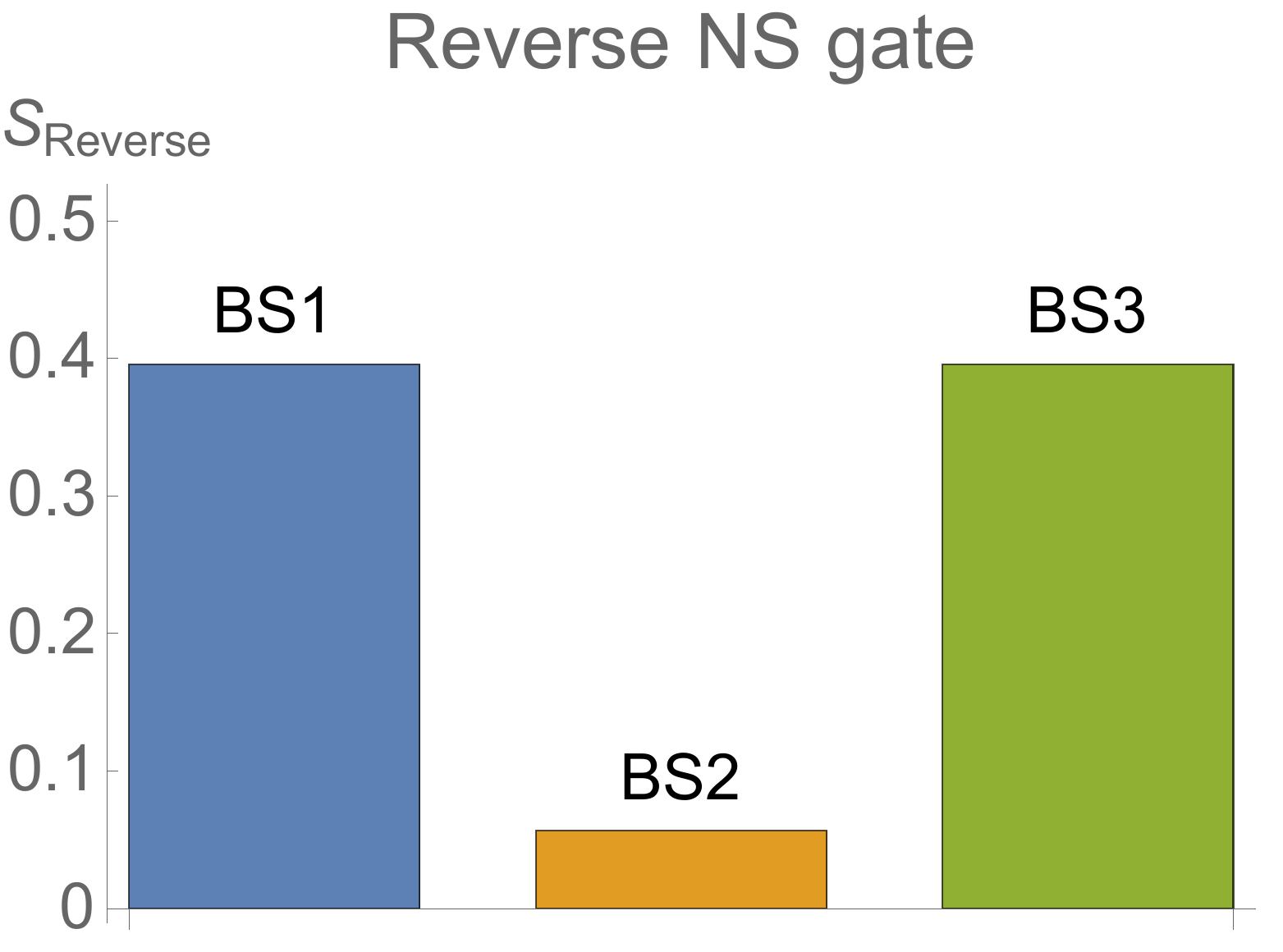} 
 \includegraphics[width=4.5cm]{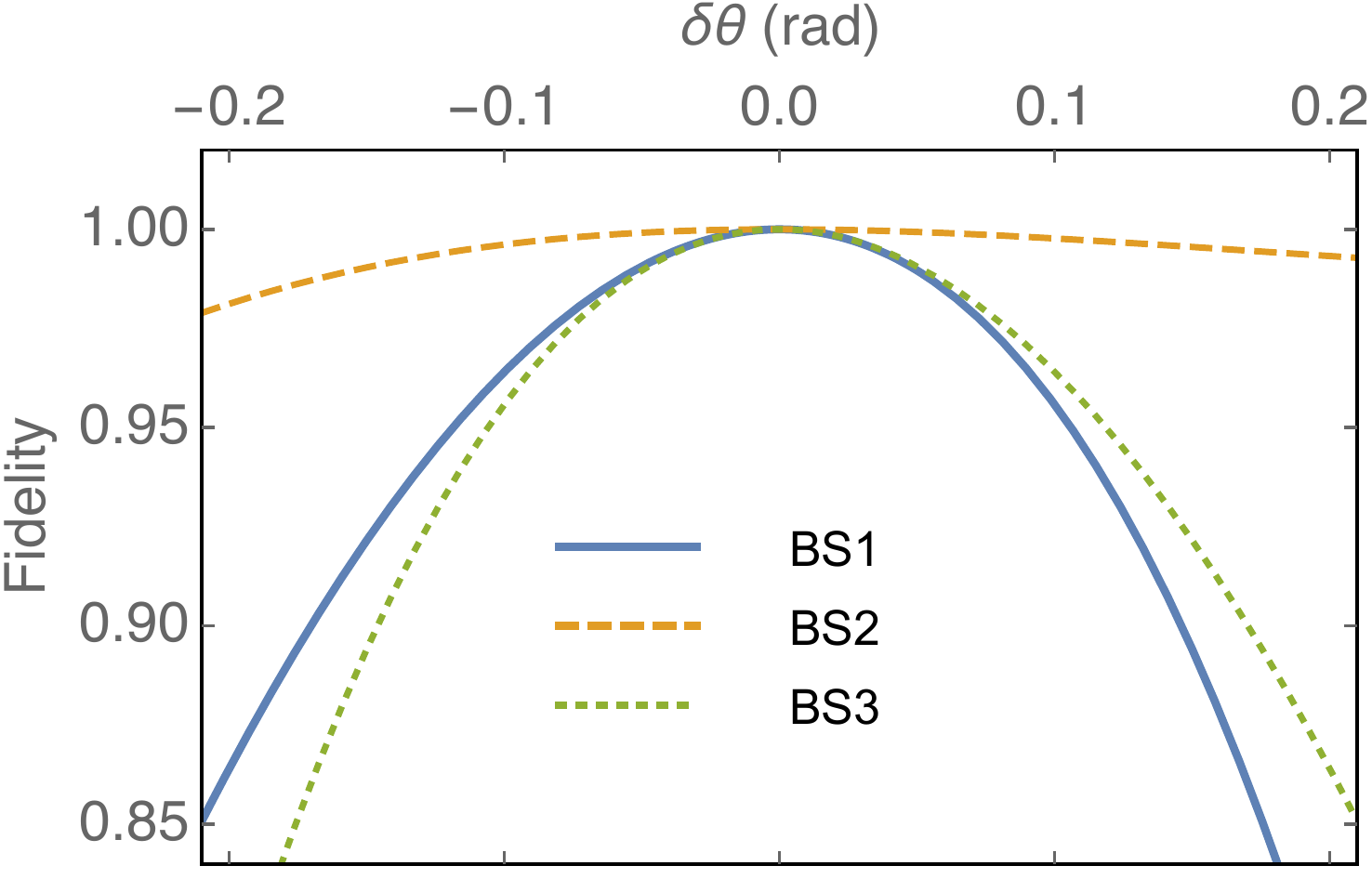} 
 \caption{Sensitivities ${S}_{\rm KLM}$ and ${S}_{\rm Reverse}$ of the beam splitters in the KLM NS gate (top) and the Reverse NS gate (bottom), averaged over all input states. The units along the vertical axis are technically rad$^{-2}$, but are not important for the comparison of similar components in a gate. These results are consistent with the detailed analysis in Ref.~\cite{Crickmore16}.}
 \label{fig:ns_sensitivity}
\end{figure}

For both the KLM and Reverse NS gate we choose the following description of the beam splitter operation acting on modes $a_1$ and $a_2$:
\begin{align}
 \begin{pmatrix}
  \hat{b}_1 \\ \hat{b}_2
 \end{pmatrix}
 =
 \begin{pmatrix}
   \cos\theta & \sin\theta \\ -\sin\theta & \cos\theta
 \end{pmatrix} 
 \begin{pmatrix}
  \hat{a}_1 \\ \hat{a}_2
 \end{pmatrix} ,
\end{align}
where hats denote mode operators and $\theta$ is the defining parameter of the beam splitter. For the KLM NS gate, the three beam splitter parameters $\theta_j$ are 
\begin{align}\label{eq:nsklmtheta}
 \theta_1 & = \arccos \left( \frac{1}{4-2\sqrt{2}} \right) \, , \cr 
 \theta_2 & = \arccos \left( 3-2\sqrt{2} \right) \, ,\cr
 \theta_3 & = -\theta_1\, ,
\end{align}
whereas for the Reverse NS gate, the three beam splitter parameters $\xi_j$ are
\begin{align}
 \xi_1 & = \arctan \left( \sqrt[4]{8} \right) \, , \cr 
 \xi_2 & = \pi - \arctan \left( \frac{\sqrt{16\sqrt{2}-13}}{7} \right) \,, \cr
 \xi_3 & = -\xi_1\, .
\end{align}
Both implementations use the same number of auxiliary photons and detections. 

We calculate the sensitivity of the two NS gates to the various beam splitters. The results are shown in Fig.~\ref{fig:ns_sensitivity}. Clearly, the sensitivity to variations in \textsf{BS2} in the KLM NS gate is much greater than the sensitivity to variations in \textsf{BS1} and \textsf{BS3}. This is reflected in the average fidelity of the output state of the KLM NS gate. By contrast, the sensitivity to variations in \textsf{BS2} in the Reverse NS gate is much smaller than the sensitivity to variations in \textsf{BS1} and \textsf{BS3}. Again, this is borne out in the average fidelities of the output state of the Reverse NS gate.

\subsection{Enhanced linear optical Bell state detectors}\label{sec:Bell}\noindent
Optical Bell measurements are an important tool in optical quantum information processing. They are used in a variety of applications, including teleportation \cite{bennett93,bouwmeester97,kok00}, optical quantum computing \cite{Browne05}, and quantum repeater proposals \cite{briegel98,Dur99,Kok03b}. Originally, the optical Bell detector was introduced by Weinfurter \cite{Weinfurter94} and Braunstein and Mann \cite{Braunstein95b}, and both schemes have a success probability of one half. In particular, these Bell detectors are capable of identifying the Bell states 
\begin{align}
 \ket{\smash{\Psi^\pm}} \equiv  \frac{\ket{H,V} \pm \ket{V,H}}{\sqrt{2}}\, , 
\end{align}
while they are completely incapable of distinguishing between the Bell states
\begin{align}
 \ket{\smash{\Phi^\pm}} \equiv  \frac{\ket{H,H} \pm \ket{V,V}}{\sqrt{2}}\, ,
\end{align}
where $\ket{H}$ and $\ket{V}$ denote horizontally and vertically polarised photons, respectively. It was proved by Vaidman and Yoran \cite{Vaidman99} and L\"utkenhaus, Calsamiglia, and Suominen \cite{Lutkenhaus99} that optical Bell detectors without auxiliary photons have an upper bound of one half on the success probability. This severely increases the overhead of any practical application relying on these Bell detectors, since provisions must be made to ensure a failed Bell measurement does not negatively affect the operation of the quantum device (e.g., see the solution provided by the original proposal for linear optical quantum computing by Knill, Laflamme, and Milburn \cite{klm01}).

\begin{figure}[t!]
\centering
 \includegraphics[width=7.5cm]{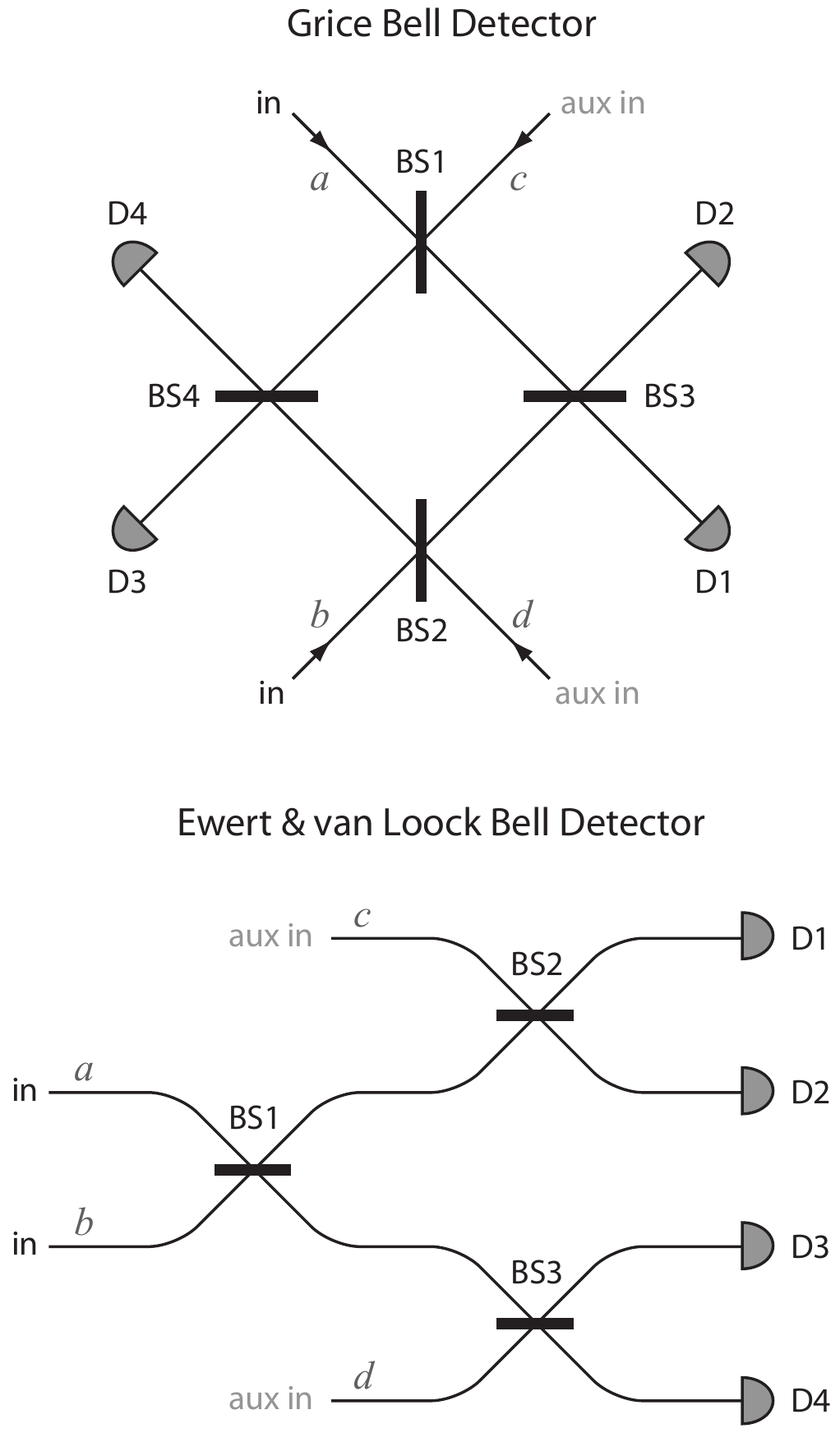}
 \caption{The entanglement-assisted Bell detection circuits of Grice (top) and Ewert \& van Loock (bottom). The Grice detector takes as input two photonic polarisation qubits and a polarisation Bell state $(\ket{H,H}+\ket{V,V})/\sqrt{2}$ in the auxiliary input. Every mode carries a polarisation degree of freedom. The Ewert \& van Loock Bell detector operates on dual-rail photonic qubits with auxiliary input states $(\ket{2,0}+\ket{0,2})/\sqrt{2}$, but here we translated it to a polarisation implementation: each mode carries a polarisation degree of freedom. In this paper we consider only the sensitivity of the Bell detection circuits to beam splitter variations.}
 \label{fig:bellcircuits}
\end{figure}

A modification of the optical Bell detector was proposed by Grice \cite{Grice11}, and Ewert \& Van Loock \cite{Ewert14}, who suggested employing auxiliary photons to help distinguish between the remaining Bell states $\ket{\smash{\Phi^\pm}}$. They showed that using one or two photon pairs increases the success probability to three quarters, and more generally, the use of $n$ photons leads to a success probability of 
\begin{align}\nonumber
 p_{\rm Grice}(n) = 1 - \frac{1}{n+2} \quad\text{and}\quad  p_{\rm EvL}(n) = 1 - \frac{1}{2^{n/2}}\, .
\end{align}
The circuits for the Grice and Ewert \& Van Loock Bell detectors using two and four auxiliary photons, respectively, is given in Fig.~\ref{fig:bellcircuits}. To obtain a success probability of three quarters, the Grice circuit requires a two-photon input state of the form
\begin{align}
 \ket{\smash{\Phi^+}} = \frac{\ket{H,H}+\ket{V,V}}{\sqrt{2}}\, ,
\end{align}
while the Ewert \& van Loock circuit requires two two-photon input states in modes $c$ and $d$ of the form
\begin{align}
 \ket{\Upsilon} = \frac{\ket{2H}+\ket{2V}}{\sqrt{2}}\, .
\end{align}
Note that while the Ewert \& van Loock circuit requires twice as many auxiliary photons to achieve the same success probability of three quarters as Grice's circuit, for higher success probabilities the Ewert \& van Loock family of circuits is more efficient.

\begin{figure}[t!]
\centering
 \includegraphics[width=4cm]{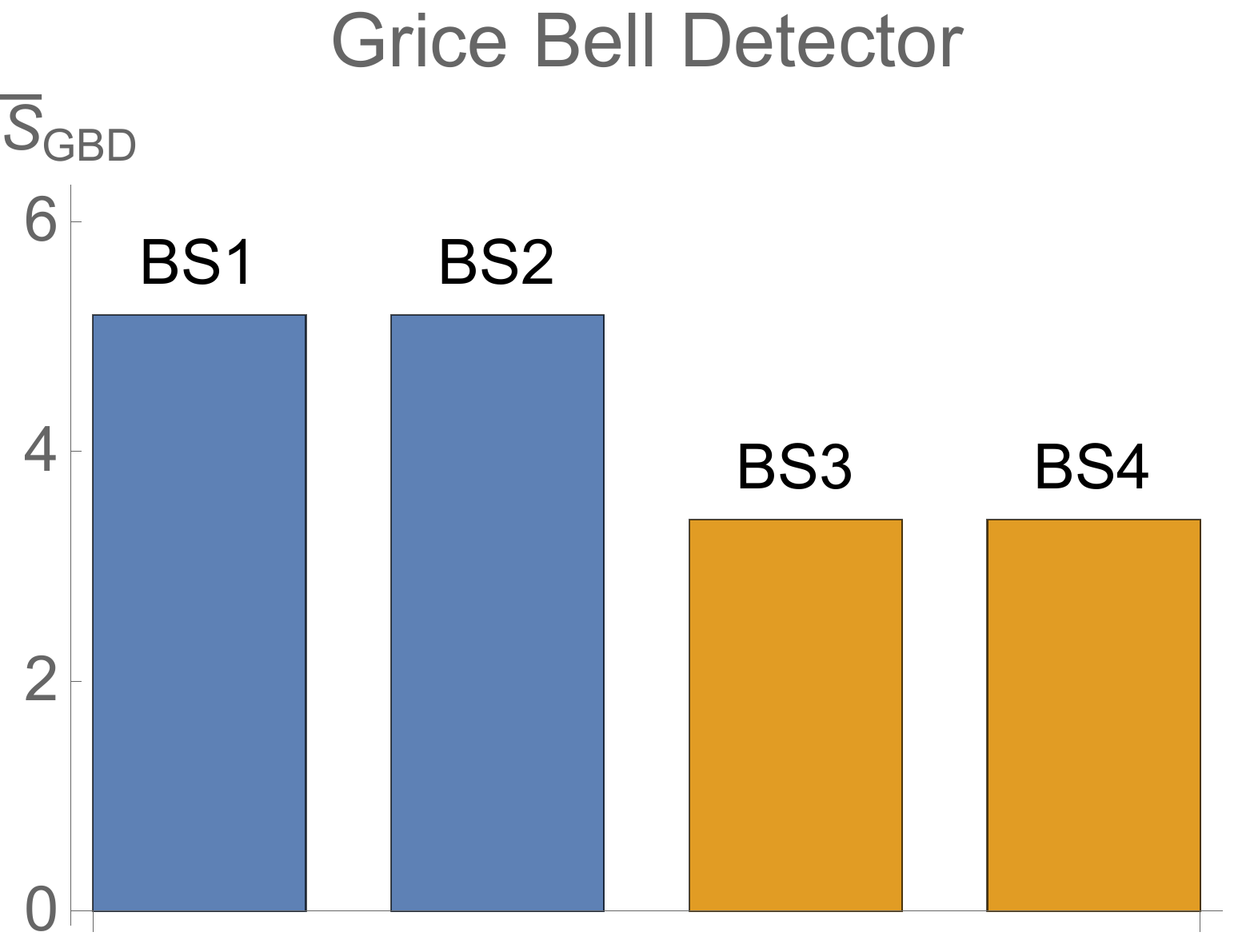}
 \includegraphics[width=4.5cm]{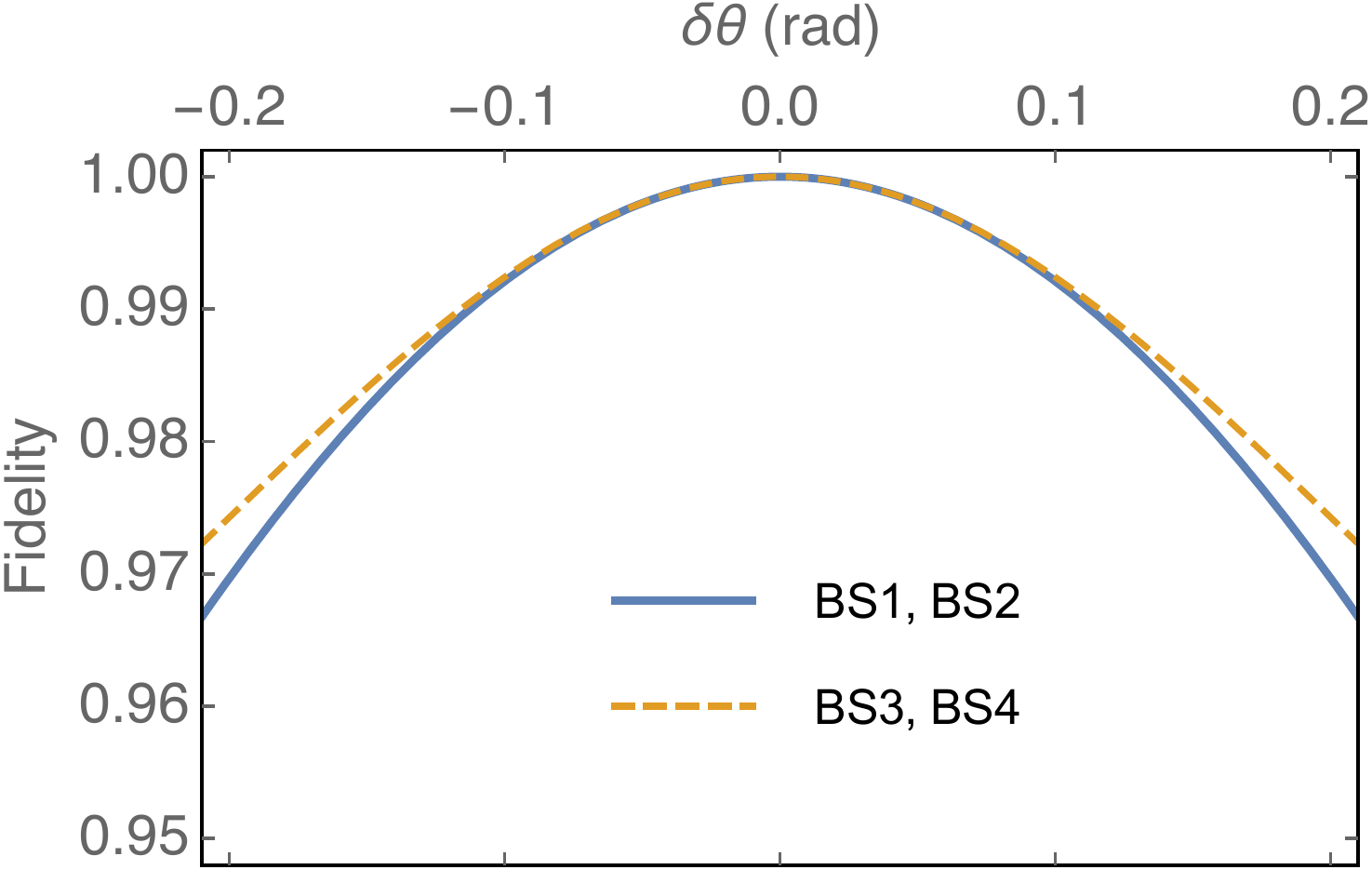} 
 \caption{The beam splitter sensitivities of the Bell detection circuits of Grice, compared with the fidelities for the different beam splitter variations. Beam splitters BS1 and BS2 are entirely equivalent, as are BS3 and BS4. The sensitivities for BS1 and BS2 are 5.2, while the sensitivities for BS1 and BS2 are 3.4. The fidelities (right) reflect this sensitivity.}
 \label{fig:gricecircuits_sensitivity}
\end{figure}

We calculate the sensitivity of the beam splitters in the Grice circuit, shown in Fig.~\ref{fig:gricecircuits_sensitivity}. We choose as input to the quantum measurement device in Fig.~\ref{fig:device}b the state
\begin{align}\label{eq:input}
 \ket{\smash{\Psi_{\rm in}}} = ~ & \frac{1}{2} \ket{\smash{\Phi^+,\Phi^+}} +  \frac{1}{2}\ket{\smash{\Phi^-,\Phi^-}} +  \frac{1}{2}\ket{\smash{\Psi^+,\Psi^+}} \cr & +  \frac{1}{2}\ket{\smash{\Psi^-,\Psi^-}}\, ,
\end{align}
and we calculate the fidelity of the output state of the device with the expected Bell state due to the measurement outcome. The first two beam splitters that the input photons encounter (\textsf{BS1} and \textsf{BS2}) exhibit a significantly lower sensitivity than the last two beam splitters (\textsf{BS3} and \textsf{BS4}). This is corroborated by the fidelity of the output state.

\begin{figure}[t!]
\centering
 \includegraphics[width=4cm]{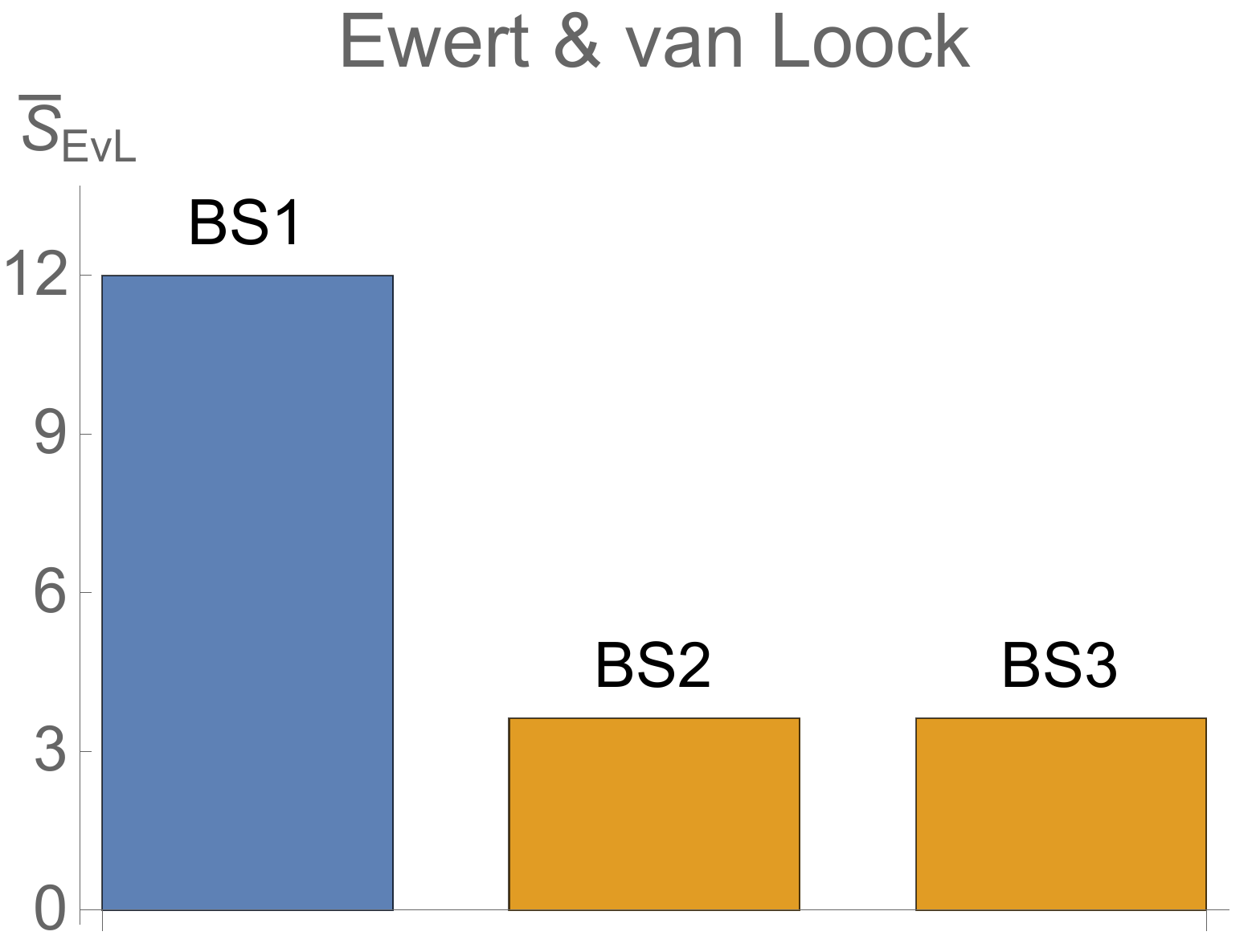}
 \includegraphics[width=4.5cm]{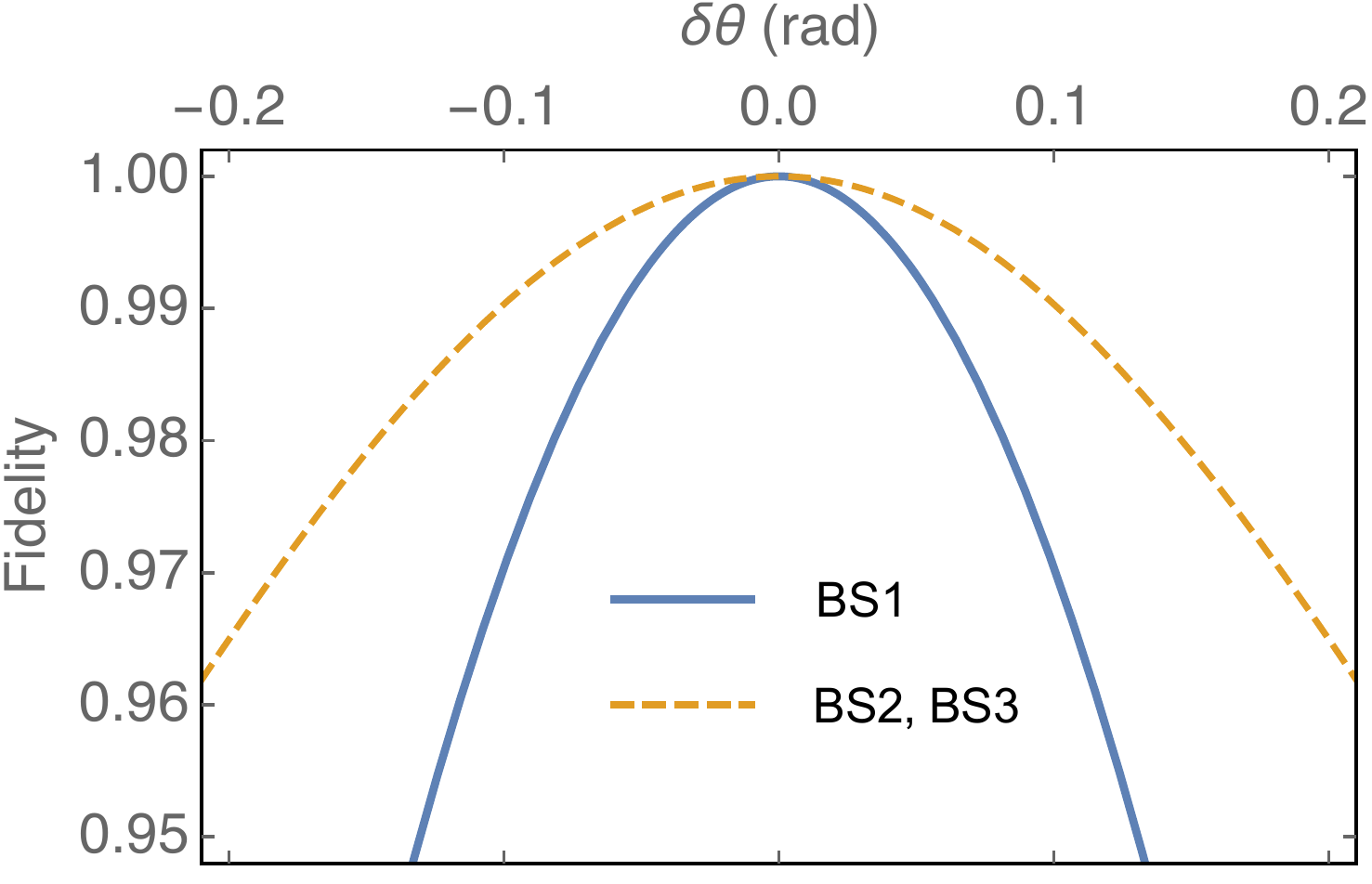} 
 \caption{The beam splitter sensitivities of the Bell detection circuits of Ewert \& van Loock, compared with the fidelities for the different beam splitter variations. Beam splitters BS2 and BS3 are equivalent. The sensitivity for BS1 is 12, while the sensitivities for BS2 and BS3 are 3.6. Again, the fidelities (right) reflect this sensitivity. In particular, the large sensitivity of the device to variations in BS1 is clear in the fidelity plot.}
 \label{fig:ewertcircuits_sensitivity}
\end{figure}

For the Ewert \& van Loock circuit we perform similar calculations, with the same input state in Eq.~(\ref{eq:input}), and we find that it is the first beam splitter (\textsf{BS1}) that exhibits the greatest sensitivity. As expected, due to the symmetry of the circuit, beamsplitters \textsf{BS2} and \textsf{BS3} have the same sensitivity. The fidelity plots again confirm the sensitivities (see Fig.~\ref{fig:ewertcircuits_sensitivity}).

It is tempting to make a judgement, based on the above analysis, which NS gate or Bell detector is more suitable for implementation. However, in our examples we have considered only the sensitivities of the beam splitters, and we have not included variations in path lengths (or phases), auxiliary input states, or detector imperfections. These must all be taken into account before a value judgement can be made about a particular quantum device or gate implementation. However, it is already clear that more resources (i.e., time spent in alignment, or money spent on high-quality components) should be devoted to \textsf{BS1} and \textsf{BS2} in the Grice circuit, and, more dramatically, to \textsf{BS1} in the Ewert \& van Loock circuit.

\section{Device analysis based on the Sensitivity}\label{sec:design}\noindent
The discussion has so far been restricted to sensitivities of individual components. However, quantum devices, including those in the previous section, typically consist of multiple components, and we would like to have some sense of which components are more critical than others to the device's operation. In general it will be hard to compare sensitivities of different types of components. For example, how should we compare the numerical values for the sensitivity of the device to a beam splitter reflectivity (dimensionless) with the sensitivity to a path length variation (units of length)? To solve this problem, we borrow the concept of cost functions from estimation theory. It will also provide us with a single metric that allows us to choose between different implementations of the same quantum device. In this section we will introduce the concept of cost functions, show how they apply to quantum device analysis, and give examples for the NS gates and Bell detectors. 

\subsection{Cost functions and their construction}\noindent
Multiple components in a device will lead to a multi-parameter estimation problem in our device analysis. Let the output state $\ket{\psi_{\rm out}}$ of a device (or $\ket{\smash{\psi_{\rm out}^{(m)}}}$ for measurement devices) depend on an array of parameters corresponding to the different device components:
\begin{align}
 \bm\theta = (\theta_1,\ldots,\theta_M)\, ,
\end{align}
where $M$ denotes the number of different components in the quantum device. In quantum metrology, for each output state $\ket{\smash{\psi_{\rm out}}}$ we assign a quantum Fisher information \emph{matrix}---defined on the parameter space of $\bm\theta$---that can be described by
\begin{align}\label{eq:eouhgdfk}
 [I_Q]_{jk} = ~ & 4\re\left[ \Braket{\partial_j{\psi}_{\rm out}|\partial_k{\psi}_{\rm out}} \right] \cr & - 4\re\left[\Braket{\partial_j\psi_{\rm out}|{\psi}_{\rm out}}\braket{\psi_{\rm out}|\partial_k{\psi}_{\rm out}}\right]\, ,
\end{align}
where $\partial_j$ and $\partial_k$ are derivatives with respect to $\theta_j$ and $\theta_k$, respectively. Calculating the derivatives as before, we obtain the matrix elements
\begin{align}
 [I_Q]_{jk} = ~& \frac{4}{p}\,\re [  \Bra{\psi_{\rm in},A} V_j^\dagger u_j^\dagger H_j W_j^\dagger (\identity-\ket{\psi_{\rm out}}\bra{\psi_{\rm out}})  \cr
 & \qquad\otimes \Pi_D\, W_k H_k u_k V_k \ket{\psi_{\rm in},A}] \, .
\end{align}
Once we have a quantum Fisher information matrix, a bound on the covariance matrix of $\bm\theta$ can be defined:
\begin{align}\label{eq:qcrbmult}
 \cov(\bm\theta) \geq I_Q^{-1}\, .
\end{align}
This is the famous quantum Cram\'er-Rao bound for multiple parameters \cite{Helstrom73}, defined in the sense that $\cov(\bm\theta) - \smash{I_Q^{-1}}$ is a positive definite matrix. The bound is tight if and only if the generators $H_j$ are co-measurable \cite{Ragy16}. The smaller the covariances, the better we can estimate variations in $\bm\theta$ from the output state, and the more sensitive the implementation is to variations in $\bm\theta$. The Sensitivity, which in Eq.~(\ref{eq:sensitivity}) was proportional to the quantum Fisher information, now becomes a Sensitivity matrix. 

In order to compare the sensitivities to different components $\theta_j$ and $\theta_k$, we need physically motived unit scales $\Delta\theta_j$ and $\Delta\theta_k$. For example, suppose that the beam splitter \textsf{BS2} in the KLM NS gate can be manufactured to much higher precision than \textsf{BS1} and  \textsf{BS3}, perhaps due to the different values of $\theta_j$ in Eq.~(\ref{eq:nsklmtheta}). Then a natural choice for the unit scale is the manufacturing tolerance $\Delta\theta_j$. Consequently, even though $S_2$ is larger than $S_1$ and $S_3$ (see Fig.~\ref{fig:ns_sensitivity}), we could be in the position that the dimensionless product of the tolerance-squared $(\Delta\theta_j)^2$  and the sensitivity $S_j$ indicates that \textsf{BS2} will have a lower impact on the device performance than \textsf{BS1} and  \textsf{BS3}:
\begin{align}
 {(\Delta\theta_2)^2} {S_2} < {(\Delta\theta_1)^2} {S_1} = {(\Delta\theta_3)^2}{S_3}\, .
\end{align}
The cost function for a single parameter can then be given as $(\Delta\theta_j)^2$. In the context of covariance bounds, a real, symmetric, positive semi-definite cost matrix $R$ is introduced such that Eq.~(\ref{eq:qcrbmult}) becomes the simple inequality 
\begin{align}\label{eq:bhu043we}
 \Tr{R \cov(\bm\theta)} \geq \Tr{R\, I_Q^{-1}}\, .
\end{align}
For our purpose of quantum device analysis, we are interested in minimising the sensitivity, rather than minimising the covariance of $\bm\theta$. We can achieve this by letting $R = \bm\Sigma(\bm\theta)$, where $\bm\Sigma(\bm\theta)$ is the covariance matrix associated with the fluctuations due to manufacturing tolerances (which is different from $\cov(\bm\theta)$ in the Cram\'er-Rao bound). In the case where components exhibit stochastic fluctuations of the parameter, the diagonal elements of $\bm\Sigma(\bm\theta)$ are the variances $(\Delta\theta_j)^2$ of the fluctuations. The dimensionless measure for the total sensitivity is then
\begin{align}
 \tr{R S} = \sum_{j,k=1}^M \bm\Sigma(\bm\theta)_{jk}\, S_{kj}\, .
\end{align}
When a component is subject to both manufacturing tolerances and stochastic fluctuations, we may combine their effects according to the standard rule
\begin{align}
 (\Delta\theta_j)^2 = (\Delta_{\rm man}\theta_j)^2 + (\Delta_{\rm st}\theta_j)^2\, ,
\end{align}
where $\Delta_{\rm man}$ and $\Delta_{\rm st}$ denote manufacturing and stochastic errors, respectively. This formula can easily be extended to include more types of noise in the components, including correlated noise. In addition, we can include the economical cost for each component as a multiplier. Such a choice will favour devices with fewer components, \emph{ceteris paribus}. 

Now consider two quantum device implementations, $\mathscr{I}_1$ and $\mathscr{I}_2$. We can calculate the sensitivity matrices $S(\mathscr{I}_1)$ and $S(\mathscr{I}_2)$ for these implementations according to Eq.~(\ref{eq:eouhgdfk}) and $S = pI_Q$. Given two cost functions $R_1$ and $R_2$, we say that implementation $\mathscr{I}_1$ is better than implementation $\mathscr{I}_2$ if 
\begin{align}\label{eq:decision}
 \Tr{R_1\, S(\mathscr{I}_1)} < \Tr{R_2\, S(\mathscr{I}_2)}.
\end{align}
Note that the decision criterion in Eq.~(\ref{eq:decision}) depends on the choice of cost functions. For example, for the NS gate implementation we can choose values of $(\Delta\theta_1,\Delta\theta_2,\Delta\theta_3)$ and $(\Delta\xi_1,\Delta\xi_2,\Delta\xi_3)$ such that either the KLM NS gate or the Reverse NS gate is a easier to implement experimentally. The practically achievable tolerances are, not surprisingly, an integral part of the device analysis.

In general, we are forced to pick two different cost functions $R_1$ and $R_2$ when the components of the two implementations differ, and we must make sure that $R_1$ and $R_2$ are constructed using the same criteria. In other words, the beam splitter variations in $\mathscr{I}_1$ should be constructed according to the same physical principles as the beam splitter variations in $\mathscr{I}_2$, even though they may not result in identical numerical values (c.f., our example of the KLM and Reverse NS gates above). This gives a natural sense in which devices with more or harder to fabricate components are likely to perform worse than simpler, easier to fabricate devices. When all components (of the same type) have identical manufacturing tolerances and are independent of the other components, the cost function may be chosen as the identity matrix.  

Finally, one may ask what the statistical interpretation of Eq.~(\ref{eq:bhu043we}) means for the task of differentiating between proposed implementations of a quantum device, and hence how we should interpret Eq.~(\ref{eq:decision}). When the Cram\'er-Rao bound in Eq.~(\ref{eq:bhu043we}) is achievable, we can extract the full quantum Fisher information's worth from measurements on the output state $\ket{\psi_{\rm out}}$ or $\ket{\smash{\psi_{\rm out}^{(m)}}}$. The sensitivity therefore immediately leads to observable effects. On the other hand, the multi-parameter Cram\'er-Rao bound cannot be saturated in general, and this leads to the question whether the decision condition in Eq.~(\ref{eq:decision}) should be modified (e.g., along the lines of Ref.~\cite{Holevo82}, using right-logarithmic derivatives). However, we should remember that we do not actually wish to estimate the parameters $\bm\theta$, but merely seek a measure of sensitivity for the \emph{quantum state} (which will typically be used in further processing) given a cost matrix $R$. This is exactly what the multi-parameter quantum Fisher information---and therefore the sensitivity---provides.

\subsection{Example of cost functions for Bell detectors}\noindent
We can make a simple comparison between the Grice Bell detector and the Ewert \& van Loock Bell detector, where we take into account only the cost of beam splitter variations. While the Grice Bell detector employs more beam splitters, the beam splitters in the Ewert \& van Loock Bell detector have a higher sensitivity. Comparing the implementations with a simple identity matrix cost function $R = \identity$ (since we have no reason to believe that there is a different cost or tolerance for these beam splitters), we find that the Ewert \& van Loock Bell detector has an overall sensitivity of $\smash{\tr{S(\mathscr{I}_{\rm EvL})}} = 19.25$, and the Grice Bell detector has an overall beam splitter sensitivity of $\smash{\tr{S(\mathscr{I}_{\rm Grice})}} = 17.19$. While the difference is not large, this shows that when we ignore path length differences and other imperfections, the design with lower sensitivities to the beam splitter variations is in this case marginally preferable to the design with fewer beam splitters. 

When a device has more components than there are degrees of freedom in the output state, the quantum Fisher information matrix may become singular. In that case we cannot evaluate Eq.~(\ref{eq:decision}), and the method presented here does not provide a value for the overall sensitivity (and this device implementation may allow for significant simplifications!). However, our method will still be able to provide valuable information about the relative importance of variations in the constituent components, and even when the sensitivity matrix is not singular, a full device analysis must always include the consideration of the individual elements. This will inform us which components will give the greatest benefits in the precision of the device when extra resources are spent on improvements.

\section{Relation to the Diamond Distance}\label{sec:diamond}\noindent
We have compared the metrological approach to quantum device characterisation to the average gate fidelity. However, another important metric for judging the quality of a quantum device is the diamond distance. Here we briefly review some key properties of the diamond distance and explore how it can be used in device analysis. 

The diamond distance is a unitarily invariant quantity for measuring the distance between two general quantum channels \cite{Kitaev97}. In the case of quantum information processors, the diamond distance tells us how much an actual gate transformation deviates from the intended gate transformation, and it is closely related to the error rate of the gate \cite{Sanders16}. Given the Schatten 1-norm $\norm[1]{\cdot}$, the diamond norm of a quantum channel $\mathscr{E}$ is defined as \cite{Kitaev97}
\begin{align}
 \norm[$\diamond$]{\mathscr{E}} \equiv \sup \left\{ \norm[1]{(\mathscr{E}\otimes\identity)(\rho)} ; \norm[1]{\rho} \leq 1 \right\}\, ,
\end{align}
which leads to a diamond distance between two channels $\mathscr{E}$ and $\mathscr{E}'$ given by
\begin{align}
 d_\diamond (\mathscr{E},\mathscr{E}') \equiv \frac12 \Norm[$\diamond$]{\mathscr{E}-\mathscr{E}'}\, .
\end{align}
It was shown in Refs.~\cite{Wang13,Wang18} that the diamond distance for unitary channels $u_j$ and $u_j'$ is upper bounded by the operator norm
\begin{align}\label{eq:vnuw90eofsdj}
 d_\diamond (u_j,u_j') \leq \norm{u_j-u_j'}\, .
\end{align}
For the types of unitary transformations discussed here, with $u_j = \exp(-i\theta_j H_j)$ and $u_j' = \exp(-i\theta_j' H_j)$, and only small difference between the parameters $\abs{\theta_j - \theta_j'}$, the diamond norm is very close to the operator norm $\norm{u_j - u_j'}$ \cite{Campbell17,Campbell18}. To see this, we first define the ground state $\ket{m}$ and the maximum eigenvalue state $\ket{M}$ of $H_j$. Without loss of generality (by fixing a global phase) the eigenvalues of $\ket{m}$ and $\ket{M}$ are $-\mu$ and $+\mu$, respectively ($\mu>0$). We can then evaluate the operator norm as
\begin{align}\label{eq:b739wrfbd}
 \Norm{u_j - u_j'} = \mu\Abs{\theta_j' - \theta_j} \, .
\end{align}
Next, we show that in addition to the upper bound in Eq.~(\ref{eq:vnuw90eofsdj}), we can construct a lower bound that approaches the upper bound in the limit of vanishing $\abs{\theta_j - \theta_j'}$. By definition,
\begin{align}\label{eq:bhg90worsj}
 \Norm[$\diamond$]{u_j - u_j'} \geq \Norm[1]{(u_j\otimes \identity) \rho (u_j^\dagger\otimes \identity) - (u_j'\otimes \identity) \rho ({u_j'}^\dagger\otimes \identity)}
\end{align}
for any quantum state $\rho$. Since any $\rho$ will provide a valid lower bound we are looking to choose $\rho$ judiciously, such that it maximises the lower bound. Let $\rho = \ket{\psi}\bra{\psi} \otimes \identity$, with 
\begin{align}\label{eq:bn5t0whrfks}
\ket{\psi} = \frac{\ket{m}+\ket{M}}{\sqrt{2}}\, .
\end{align} 
We then obtain \cite{Campbell18}
\begin{align}\label{eq:b902wesd}
 \Norm[$\diamond$]{u_j - u_j'} \geq 2 \mu\Abs{\theta_j' - \theta_j} .
\end{align}
Combining the upper and lower bounds in Eqs.~(\ref{eq:b739wrfbd}) and (\ref{eq:b902wesd}) to the diamond distance, and defining $\delta\theta_j = \abs{\theta_j' - \theta_j}$ (assuming $\theta_j$ is the ideal value), this leads to  
\begin{align}\label{eq:bn8049wesojidfknv}
 d_\diamond (u_j,u_j') =  \mu\, \delta\theta_j\, .
\end{align}
This is the diamond distance between $u_j$ and $u_j'$, which depends again on the deviation of $\theta_j'$ away from $\theta_j$. It can be interpreted as the maximum possible angle between two quantum states $\ket{\psi_0}$ and $\ket{\psi(\delta\theta_j)} \equiv \exp(-i\delta\theta_j H_j)\ket{\psi_0}$, where the maximisation is over the input states $\ket{\psi_0}$. Our choice for $\rho$ in Eq.~(\ref{eq:bn5t0whrfks}) achieves this maximum. 

For pure states, the angle between quantum states is known as the statistical distance between those states \cite{Wootters81}. The square of the derivative of the statistical distance with respect to the angle is equal to the quantum Fisher information for that angle. Therefore, the diamond distance for a device component---translated into a Fisher information---provides us with the largest amount of information about the component parameter $\theta_j$ that can be extracted in a measurement procedure, since the diamond distance involves a maximisation over the input states of the component. By contrast, the sensitivity presented in this paper does not involve such a maximisation, but instead uses the quantum state of the system as produced at the point just before the component $u_j$. As such, the sensitivity allows us to consider each component within the context of the device as a whole, rather than as a stand-alone device for which we seek the worst-case behaviour.  The diamond distance not distinguish between similar elements in a device, e.g., the beam splitters in the different implementations of the NS gates or the Bell detectors. The diamond distance for individual components therefore cannot be used to identify sensitivity bottlenecks in a quantum device.

Finally, note that we have assumed a maximum eigenvalue $\mu>0$ for the Hamiltonian $H_j$. In our examples involving optical modes, such Hamiltonians are typically unbounded. In our metrological approach we circumvent this problem by averaging the sensitivity over the relevant input state space. For example, the NS gate acts on at most two photons, and the averaging is performed relative to the Haar measure on the space of zero, one and two photons in the input state. For the diamond distance to be meaningful, a similar state space truncation must be employed.

\bigskip

\section{Conclusions}\label{sec:conclusions}\noindent
The construction of quantum information processing devices is a challenging technical problem, and reducing sources of errors is an essential element of it. We introduced a method for comparing different physical implementations of a quantum information processing device, including composite quantum gates and detectors, in terms of the sensitivity of the device to variations in its components. Our method is based on the amount of information about a component parameter present in the output state. This is measured by the quantum Fisher information. For the examples considered here, we show that this method is consistent with the predictions based on the fidelity of the device given variations in the components. The benefit of our method over the fidelity method is that we can collect the combined effect of variations in all components into a single overall sensitivity metric based on cost functions that match our design requirements. Furthermore, the sensitivity takes into account the context of each component in a quantum device, as opposed to the diamond distance, which captures the worst-case behaviour of each component.

For each quantum device, quantum architecture designers should consider as many different implementations as possible, and carry out a sensitivity analysis along the lines of the discussion presented here. This may be a lengthy task, and it is currently an open question whether we can construct guiding design principles that provide shortcuts to this task. Such principles will likely be highly implementation dependent.

\section*{Acknowledgements}\noindent
The authors thank Earl T.\ Campbell for valuable discussions about the diamond distance and its relation to the Fisher information.
This research was funded in part by EPSRC's Quantum Communications Hub \texttt{EP/M013472/1}. 


\begin{thebibliography}{45}%
\makeatletter
\providecommand \@ifxundefined [1]{%
 \@ifx{#1\undefined}
}%
\providecommand \@ifnum [1]{%
 \ifnum #1\expandafter \@firstoftwo
 \else \expandafter \@secondoftwo
 \fi
}%
\providecommand \@ifx [1]{%
 \ifx #1\expandafter \@firstoftwo
 \else \expandafter \@secondoftwo
 \fi
}%
\providecommand \natexlab [1]{#1}%
\providecommand \enquote  [1]{``#1''}%
\providecommand \bibnamefont  [1]{#1}%
\providecommand \bibfnamefont [1]{#1}%
\providecommand \citenamefont [1]{#1}%
\providecommand \href@noop [0]{\@secondoftwo}%
\providecommand \href [0]{\begingroup \@sanitize@url \@href}%
\providecommand \@href[1]{\@@startlink{#1}\@@href}%
\providecommand \@@href[1]{\endgroup#1\@@endlink}%
\providecommand \@sanitize@url [0]{\catcode `\\12\catcode `\$12\catcode
  `\&12\catcode `\#12\catcode `\^12\catcode `\_12\catcode `\%12\relax}%
\providecommand \@@startlink[1]{}%
\providecommand \@@endlink[0]{}%
\providecommand \url  [0]{\begingroup\@sanitize@url \@url }%
\providecommand \@url [1]{\endgroup\@href {#1}{\urlprefix }}%
\providecommand \urlprefix  [0]{URL }%
\providecommand \Eprint [0]{\href }%
\providecommand \doibase [0]{http://dx.doi.org/}%
\providecommand \selectlanguage [0]{\@gobble}%
\providecommand \bibinfo  [0]{\@secondoftwo}%
\providecommand \bibfield  [0]{\@secondoftwo}%
\providecommand \translation [1]{[#1]}%
\providecommand \BibitemOpen [0]{}%
\providecommand \bibitemStop [0]{}%
\providecommand \bibitemNoStop [0]{.\EOS\space}%
\providecommand \EOS [0]{\spacefactor3000\relax}%
\providecommand \BibitemShut  [1]{\csname bibitem#1\endcsname}%
\let\auto@bib@innerbib\@empty
\bibitem [{\citenamefont {Raussendorf}\ \emph {et~al.}(2007)\citenamefont
  {Raussendorf}, \citenamefont {Harrington},\ and\ \citenamefont
  {Goyal}}]{Raussendorf07}%
  \BibitemOpen
  \bibfield  {author} {\bibinfo {author} {\bibfnamefont {R.}~\bibnamefont
  {Raussendorf}}, \bibinfo {author} {\bibfnamefont {J.}~\bibnamefont
  {Harrington}}, \ and\ \bibinfo {author} {\bibfnamefont {K.}~\bibnamefont
  {Goyal}},\ }\href {\doibase 10.1088/1367-2630/9/6/199} {\bibfield  {journal}
  {\bibinfo  {journal} {New J. Phys.}\ }\textbf {\bibinfo {volume} {9}},\
  \bibinfo {pages} {199} (\bibinfo {year} {2007})}\BibitemShut {NoStop}%
\bibitem [{\citenamefont {Demkowicz-Dobrzanski}\ \emph
  {et~al.}(2012)\citenamefont {Demkowicz-Dobrzanski}, \citenamefont
  {Kolodynski},\ and\ \citenamefont {Guta}}]{Dobrzanski12}%
  \BibitemOpen
  \bibfield  {author} {\bibinfo {author} {\bibfnamefont {R.}~\bibnamefont
  {Demkowicz-Dobrzanski}}, \bibinfo {author} {\bibfnamefont {J.}~\bibnamefont
  {Kolodynski}}, \ and\ \bibinfo {author} {\bibfnamefont {M.}~\bibnamefont
  {Guta}},\ }\href {\doibase 10.1038/ncomms2067} {\bibfield  {journal}
  {\bibinfo  {journal} {Nature Commun.}\ }\textbf {\bibinfo {volume} {3}},\
  \bibinfo {pages} {1063} (\bibinfo {year} {2012})}\BibitemShut {NoStop}%
\bibitem [{\citenamefont {Uhlmann}(1976)}]{Uhlmann76}%
  \BibitemOpen
  \bibfield  {author} {\bibinfo {author} {\bibfnamefont {A.~A.}\ \bibnamefont
  {Uhlmann}},\ }\href {\doibase 10.1016/0034-4877(76)90060-4} {\bibfield
  {journal} {\bibinfo  {journal} {Rep. Math. Phys.}\ }\textbf {\bibinfo
  {volume} {9}},\ \bibinfo {pages} {273} (\bibinfo {year} {1976})}\BibitemShut
  {NoStop}%
\bibitem [{\citenamefont {Myerson}\ \emph {et~al.}(2008)\citenamefont
  {Myerson}, \citenamefont {Szwer}, \citenamefont {Webster}, \citenamefont
  {Allcock}, \citenamefont {Curtis}, \citenamefont {Imreh}, \citenamefont
  {Sherman}, \citenamefont {Stacey}, \citenamefont {Steane},\ and\
  \citenamefont {Lucas}}]{Lucas08}%
  \BibitemOpen
  \bibfield  {author} {\bibinfo {author} {\bibfnamefont {A.~H.}\ \bibnamefont
  {Myerson}}, \bibinfo {author} {\bibfnamefont {D.~J.}\ \bibnamefont {Szwer}},
  \bibinfo {author} {\bibfnamefont {S.~C.}\ \bibnamefont {Webster}}, \bibinfo
  {author} {\bibfnamefont {D.~T.~C.}\ \bibnamefont {Allcock}}, \bibinfo
  {author} {\bibfnamefont {M.~J.}\ \bibnamefont {Curtis}}, \bibinfo {author}
  {\bibfnamefont {G.}~\bibnamefont {Imreh}}, \bibinfo {author} {\bibfnamefont
  {J.~A.}\ \bibnamefont {Sherman}}, \bibinfo {author} {\bibfnamefont {D.~N.}\
  \bibnamefont {Stacey}}, \bibinfo {author} {\bibfnamefont {A.~M.}\
  \bibnamefont {Steane}}, \ and\ \bibinfo {author} {\bibfnamefont {D.~M.}\
  \bibnamefont {Lucas}},\ }\href {\doibase 10.1103/PhysRevLett.100.200502}
  {\bibfield  {journal} {\bibinfo  {journal} {Phys. Rev. Lett.}\ }\textbf
  {\bibinfo {volume} {100}},\ \bibinfo {pages} {171} (\bibinfo {year}
  {2008})}\BibitemShut {NoStop}%
\bibitem [{\citenamefont {Ballance}\ \emph {et~al.}(2016)\citenamefont
  {Ballance}, \citenamefont {Harty}, \citenamefont {Linke}, \citenamefont
  {Sepiol},\ and\ \citenamefont {Lucas}}]{Lucas16}%
  \BibitemOpen
  \bibfield  {author} {\bibinfo {author} {\bibfnamefont {C.~J.}\ \bibnamefont
  {Ballance}}, \bibinfo {author} {\bibfnamefont {T.~P.}\ \bibnamefont {Harty}},
  \bibinfo {author} {\bibfnamefont {N.~M.}\ \bibnamefont {Linke}}, \bibinfo
  {author} {\bibfnamefont {M.~A.}\ \bibnamefont {Sepiol}}, \ and\ \bibinfo
  {author} {\bibfnamefont {D.~M.}\ \bibnamefont {Lucas}},\ }\href {\doibase
  10.1103/PhysRevLett.117.060504} {\bibfield  {journal} {\bibinfo  {journal}
  {Phys. Rev. Lett.}\ }\textbf {\bibinfo {volume} {117}},\ \bibinfo {pages}
  {060504} (\bibinfo {year} {2016})}\BibitemShut {NoStop}%
\bibitem [{\citenamefont {Ghosh}\ \emph {et~al.}(2013)\citenamefont {Ghosh},
  \citenamefont {Galiautdinov}, \citenamefont {Zhou},\ and\ \citenamefont
  {Korotkov}}]{Ghosh13}%
  \BibitemOpen
  \bibfield  {author} {\bibinfo {author} {\bibfnamefont {J.}~\bibnamefont
  {Ghosh}}, \bibinfo {author} {\bibfnamefont {A.}~\bibnamefont {Galiautdinov}},
  \bibinfo {author} {\bibfnamefont {Z.}~\bibnamefont {Zhou}}, \ and\ \bibinfo
  {author} {\bibfnamefont {A.~N.}\ \bibnamefont {Korotkov}},\ }\href {\doibase
  10.1103/PhysRevA.87.022309} {\bibfield  {journal} {\bibinfo  {journal} {Phys.
  Rev. A}\ }\textbf {\bibinfo {volume} {87}},\ \bibinfo {pages} {022309}
  (\bibinfo {year} {2013})}\BibitemShut {NoStop}%
\bibitem [{\citenamefont {Pedersen}\ \emph {et~al.}(2007)\citenamefont
  {Pedersen}, \citenamefont {M{\o}ller},\ and\ \citenamefont
  {M{\o}lmer}}]{Molmer07}%
  \BibitemOpen
  \bibfield  {author} {\bibinfo {author} {\bibfnamefont {L.~H.}\ \bibnamefont
  {Pedersen}}, \bibinfo {author} {\bibfnamefont {N.~M.}\ \bibnamefont
  {M{\o}ller}}, \ and\ \bibinfo {author} {\bibfnamefont {K.}~\bibnamefont
  {M{\o}lmer}},\ }\href {\doibase 10.1016/j.physleta.2007.02.069} {\bibfield
  {journal} {\bibinfo  {journal} {Phys. Lett. A}\ }\textbf {\bibinfo {volume}
  {367}},\ \bibinfo {pages} {47} (\bibinfo {year} {2007})}\BibitemShut
  {NoStop}%
\bibitem [{\citenamefont {Cui}\ and\ \citenamefont {Wang}(2014)}]{Cui14}%
  \BibitemOpen
  \bibfield  {author} {\bibinfo {author} {\bibfnamefont {J.-X.}\ \bibnamefont
  {Cui}}\ and\ \bibinfo {author} {\bibfnamefont {Z.-D.}\ \bibnamefont {Wang}},\
  }\href {\doibase 10.1140/epjd/e2014-50354-5} {\bibfield  {journal} {\bibinfo
  {journal} {Eur. Phys. J. D}\ }\textbf {\bibinfo {volume} {68}},\ \bibinfo
  {pages} {325} (\bibinfo {year} {2014})}\BibitemShut {NoStop}%
\bibitem [{\citenamefont {Holz{\"a}pfel}\ \emph {et~al.}(2015)\citenamefont
  {Holz{\"a}pfel}, \citenamefont {Baumgratz}, \citenamefont {Cramer},\ and\
  \citenamefont {Plenio}}]{Plenio15}%
  \BibitemOpen
  \bibfield  {author} {\bibinfo {author} {\bibfnamefont {M.}~\bibnamefont
  {Holz{\"a}pfel}}, \bibinfo {author} {\bibfnamefont {T.}~\bibnamefont
  {Baumgratz}}, \bibinfo {author} {\bibfnamefont {M.}~\bibnamefont {Cramer}}, \
  and\ \bibinfo {author} {\bibfnamefont {M.~B.}\ \bibnamefont {Plenio}},\
  }\href {\doibase 10.1103/PhysRevA.91.042129} {\bibfield  {journal} {\bibinfo
  {journal} {Phys. Rev. A}\ }\textbf {\bibinfo {volume} {91}},\ \bibinfo
  {pages} {401} (\bibinfo {year} {2015})}\BibitemShut {NoStop}%
\bibitem [{\citenamefont {Kong}\ \emph {et~al.}(2017)\citenamefont {Kong},
  \citenamefont {Zong}, \citenamefont {Yang},\ and\ \citenamefont
  {Cao}}]{Kong17}%
  \BibitemOpen
  \bibfield  {author} {\bibinfo {author} {\bibfnamefont {F.-Z.}\ \bibnamefont
  {Kong}}, \bibinfo {author} {\bibfnamefont {X.-L.}\ \bibnamefont {Zong}},
  \bibinfo {author} {\bibfnamefont {M.}~\bibnamefont {Yang}}, \ and\ \bibinfo
  {author} {\bibfnamefont {Z.-L.}\ \bibnamefont {Cao}},\ }\href@noop {}
  {\bibfield  {journal} {\bibinfo  {journal} {Laser Phys. Lett.}\ }\textbf
  {\bibinfo {volume} {13}},\ \bibinfo {pages} {045201} (\bibinfo {year}
  {2017})}\BibitemShut {NoStop}%
\bibitem [{\citenamefont {Sanders}\ \emph {et~al.}(2016)\citenamefont
  {Sanders}, \citenamefont {Wallman},\ and\ \citenamefont
  {Sanders}}]{Sanders16}%
  \BibitemOpen
  \bibfield  {author} {\bibinfo {author} {\bibfnamefont {Y.~R.}\ \bibnamefont
  {Sanders}}, \bibinfo {author} {\bibfnamefont {J.~J.}\ \bibnamefont
  {Wallman}}, \ and\ \bibinfo {author} {\bibfnamefont {B.~C.}\ \bibnamefont
  {Sanders}},\ }\href {\doibase 10.1088/1367-2630/18/1/012002} {\bibfield
  {journal} {\bibinfo  {journal} {New J. Phys.}\ }\textbf {\bibinfo {volume}
  {18}} (\bibinfo {year} {2016}),\ 10.1088/1367-2630/18/1/012002}\BibitemShut
  {NoStop}%
\bibitem [{\citenamefont {Crickmore}\ \emph {et~al.}(2016)\citenamefont
  {Crickmore}, \citenamefont {Frazer}, \citenamefont {Shaw},\ and\
  \citenamefont {Kok}}]{Crickmore16}%
  \BibitemOpen
  \bibfield  {author} {\bibinfo {author} {\bibfnamefont {J.}~\bibnamefont
  {Crickmore}}, \bibinfo {author} {\bibfnamefont {J.}~\bibnamefont {Frazer}},
  \bibinfo {author} {\bibfnamefont {S.}~\bibnamefont {Shaw}}, \ and\ \bibinfo
  {author} {\bibfnamefont {P.}~\bibnamefont {Kok}},\ }\href {\doibase
  10.1103/PhysRevA.94.022326} {\bibfield  {journal} {\bibinfo  {journal} {Phys.
  Rev. A}\ }\textbf {\bibinfo {volume} {94}},\ \bibinfo {pages} {022326}
  (\bibinfo {year} {2016})}\BibitemShut {NoStop}%
\bibitem [{\citenamefont {Knill}\ \emph {et~al.}(2001)\citenamefont {Knill},
  \citenamefont {Laflamme},\ and\ \citenamefont {Milburn}}]{klm01}%
  \BibitemOpen
  \bibfield  {author} {\bibinfo {author} {\bibfnamefont {E.}~\bibnamefont
  {Knill}}, \bibinfo {author} {\bibfnamefont {R.}~\bibnamefont {Laflamme}}, \
  and\ \bibinfo {author} {\bibfnamefont {G.~J.}\ \bibnamefont {Milburn}},\
  }\href {\doibase 10.1038/35051009} {\bibfield  {journal} {\bibinfo  {journal}
  {Nature}\ }\textbf {\bibinfo {volume} {409}},\ \bibinfo {pages} {46}
  (\bibinfo {year} {2001})}\BibitemShut {NoStop}%
\bibitem [{\citenamefont {Grice}(2011)}]{Grice11}%
  \BibitemOpen
  \bibfield  {author} {\bibinfo {author} {\bibfnamefont {W.~P.}\ \bibnamefont
  {Grice}},\ }\href {\doibase 10.1103/PhysRevA.84.042331} {\bibfield  {journal}
  {\bibinfo  {journal} {Phys. Rev. A}\ }\textbf {\bibinfo {volume} {84}},\
  \bibinfo {pages} {042331} (\bibinfo {year} {2011})}\BibitemShut {NoStop}%
\bibitem [{\citenamefont {Ewert}\ and\ \citenamefont {van
  Loock}(2014)}]{Ewert14}%
  \BibitemOpen
  \bibfield  {author} {\bibinfo {author} {\bibfnamefont {F.}~\bibnamefont
  {Ewert}}\ and\ \bibinfo {author} {\bibfnamefont {P.}~\bibnamefont {van
  Loock}},\ }\href {\doibase 10.1103/PhysRevLett.113.140403} {\bibfield
  {journal} {\bibinfo  {journal} {Phys. Rev. Lett.}\ }\textbf {\bibinfo
  {volume} {113}},\ \bibinfo {pages} {140403} (\bibinfo {year}
  {2014})}\BibitemShut {NoStop}%
\bibitem [{\citenamefont {Kok}\ and\ \citenamefont
  {Braunstein}(2001)}]{kok01b}%
  \BibitemOpen
  \bibfield  {author} {\bibinfo {author} {\bibfnamefont {P.}~\bibnamefont
  {Kok}}\ and\ \bibinfo {author} {\bibfnamefont {S.~L.}\ \bibnamefont
  {Braunstein}},\ }\href {\doibase 10.1103/PhysRevA.63.033812} {\bibfield
  {journal} {\bibinfo  {journal} {Phys. Rev. A}\ }\textbf {\bibinfo {volume}
  {63}},\ \bibinfo {pages} {033812} (\bibinfo {year} {2001})}\BibitemShut
  {NoStop}%
\bibitem [{\citenamefont {Braunstein}\ \emph {et~al.}(1995)\citenamefont
  {Braunstein}, \citenamefont {Caves},\ and\ \citenamefont
  {Milburn}}]{braunstein95}%
  \BibitemOpen
  \bibfield  {author} {\bibinfo {author} {\bibfnamefont {S.~L.}\ \bibnamefont
  {Braunstein}}, \bibinfo {author} {\bibfnamefont {C.~M.}\ \bibnamefont
  {Caves}}, \ and\ \bibinfo {author} {\bibfnamefont {G.~J.}\ \bibnamefont
  {Milburn}},\ }\href {\doibase 10.1006/aphy.1996.0040} {\bibfield  {journal}
  {\bibinfo  {journal} {Ann. Phys.}\ }\textbf {\bibinfo {volume} {247}},\
  \bibinfo {pages} {135} (\bibinfo {year} {1995})}\BibitemShut {NoStop}%
\bibitem [{\citenamefont {Barrett}\ and\ \citenamefont
  {Kok}(2005)}]{Barrett05}%
  \BibitemOpen
  \bibfield  {author} {\bibinfo {author} {\bibfnamefont {S.~D.}\ \bibnamefont
  {Barrett}}\ and\ \bibinfo {author} {\bibfnamefont {P.}~\bibnamefont {Kok}},\
  }\href {\doibase 10.1103/PhysRevA.71.060310} {\bibfield  {journal} {\bibinfo
  {journal} {Phys. Rev. A}\ }\textbf {\bibinfo {volume} {71}},\ \bibinfo
  {pages} {060310} (\bibinfo {year} {2005})}\BibitemShut {NoStop}%
\bibitem [{\citenamefont {Glancy}\ \emph {et~al.}(2002)\citenamefont {Glancy},
  \citenamefont {LoSecco}, \citenamefont {Vasconcelos},\ and\ \citenamefont
  {Tanner}}]{glancy02}%
  \BibitemOpen
  \bibfield  {author} {\bibinfo {author} {\bibfnamefont {S.}~\bibnamefont
  {Glancy}}, \bibinfo {author} {\bibfnamefont {J.~M.}\ \bibnamefont {LoSecco}},
  \bibinfo {author} {\bibfnamefont {H.~M.}\ \bibnamefont {Vasconcelos}}, \ and\
  \bibinfo {author} {\bibfnamefont {C.~E.}\ \bibnamefont {Tanner}},\ }\href
  {\doibase 10.1103/PhysRevA.65.062317} {\bibfield  {journal} {\bibinfo
  {journal} {Phys. Rev. A}\ }\textbf {\bibinfo {volume} {65}},\ \bibinfo
  {pages} {062317} (\bibinfo {year} {2002})}\BibitemShut {NoStop}%
\bibitem [{\citenamefont {Ralph}\ \emph {et~al.}(2002)\citenamefont {Ralph},
  \citenamefont {Langford}, \citenamefont {Bell},\ and\ \citenamefont
  {White}}]{ralph02a}%
  \BibitemOpen
  \bibfield  {author} {\bibinfo {author} {\bibfnamefont {T.~C.}\ \bibnamefont
  {Ralph}}, \bibinfo {author} {\bibfnamefont {N.~K.}\ \bibnamefont {Langford}},
  \bibinfo {author} {\bibfnamefont {T.~B.}\ \bibnamefont {Bell}}, \ and\
  \bibinfo {author} {\bibfnamefont {A.~G.}\ \bibnamefont {White}},\ }\href
  {\doibase 10.1103/PhysRevA.65.012314} {\bibfield  {journal} {\bibinfo
  {journal} {Phys. Rev. A}\ }\textbf {\bibinfo {volume} {65}},\ \bibinfo
  {pages} {062324} (\bibinfo {year} {2002})}\BibitemShut {NoStop}%
\bibitem [{\citenamefont {Lund}\ \emph {et~al.}(2003)\citenamefont {Lund},
  \citenamefont {Bell},\ and\ \citenamefont {Ralph}}]{lund03}%
  \BibitemOpen
  \bibfield  {author} {\bibinfo {author} {\bibfnamefont {A.~P.}\ \bibnamefont
  {Lund}}, \bibinfo {author} {\bibfnamefont {T.~B.}\ \bibnamefont {Bell}}, \
  and\ \bibinfo {author} {\bibfnamefont {T.~C.}\ \bibnamefont {Ralph}},\ }\href
  {\doibase 10.1103/PhysRevA.68.022313} {\bibfield  {journal} {\bibinfo
  {journal} {Phys. Rev. A}\ }\textbf {\bibinfo {volume} {68}},\ \bibinfo
  {pages} {022313} (\bibinfo {year} {2003})}\BibitemShut {NoStop}%
\bibitem [{\citenamefont {Rohde}\ \emph
  {et~al.}(2005{\natexlab{a}})\citenamefont {Rohde}, \citenamefont {Pryde},
  \citenamefont {O'Brien},\ and\ \citenamefont {Ralph}}]{rohde05a}%
  \BibitemOpen
  \bibfield  {author} {\bibinfo {author} {\bibfnamefont {P.~P.}\ \bibnamefont
  {Rohde}}, \bibinfo {author} {\bibfnamefont {G.~J.}\ \bibnamefont {Pryde}},
  \bibinfo {author} {\bibfnamefont {J.~L.}\ \bibnamefont {O'Brien}}, \ and\
  \bibinfo {author} {\bibfnamefont {T.~C.}\ \bibnamefont {Ralph}},\ }\href
  {\doibase 10.1103/PhysRevA.72.032306} {\bibfield  {journal} {\bibinfo
  {journal} {Phys. Rev. A}\ }\textbf {\bibinfo {volume} {72}},\ \bibinfo
  {pages} {032306} (\bibinfo {year} {2005}{\natexlab{a}})}\BibitemShut
  {NoStop}%
\bibitem [{\citenamefont {Rohde}\ \emph
  {et~al.}(2005{\natexlab{b}})\citenamefont {Rohde}, \citenamefont {Ralph},\
  and\ \citenamefont {Nielsen}}]{rohde05b}%
  \BibitemOpen
  \bibfield  {author} {\bibinfo {author} {\bibfnamefont {P.~P.}\ \bibnamefont
  {Rohde}}, \bibinfo {author} {\bibfnamefont {T.~C.}\ \bibnamefont {Ralph}}, \
  and\ \bibinfo {author} {\bibfnamefont {M.~A.}\ \bibnamefont {Nielsen}},\
  }\href {\doibase 10.1103/PhysRevA.72.052332} {\bibfield  {journal} {\bibinfo
  {journal} {Phys. Rev. A}\ }\textbf {\bibinfo {volume} {72}},\ \bibinfo
  {pages} {052332} (\bibinfo {year} {2005}{\natexlab{b}})}\BibitemShut
  {NoStop}%
\bibitem [{\citenamefont {Rohde}\ and\ \citenamefont {Ralph}(2005)}]{rohde05c}%
  \BibitemOpen
  \bibfield  {author} {\bibinfo {author} {\bibfnamefont {P.~P.}\ \bibnamefont
  {Rohde}}\ and\ \bibinfo {author} {\bibfnamefont {T.~C.}\ \bibnamefont
  {Ralph}},\ }\href {\doibase 10.1103/PhysRevA.71.032320} {\bibfield  {journal}
  {\bibinfo  {journal} {Phys. Rev. A}\ }\textbf {\bibinfo {volume} {71}},\
  \bibinfo {pages} {032320} (\bibinfo {year} {2005})}\BibitemShut {NoStop}%
\bibitem [{\citenamefont {Clements}\ \emph {et~al.}(2016)\citenamefont
  {Clements}, \citenamefont {Humphreys}, \citenamefont {Metcalf}, \citenamefont
  {Kolthammer},\ and\ \citenamefont {Walmsley}}]{Clements16}%
  \BibitemOpen
  \bibfield  {author} {\bibinfo {author} {\bibfnamefont {W.~R.}\ \bibnamefont
  {Clements}}, \bibinfo {author} {\bibfnamefont {P.~C.}\ \bibnamefont
  {Humphreys}}, \bibinfo {author} {\bibfnamefont {B.~J.}\ \bibnamefont
  {Metcalf}}, \bibinfo {author} {\bibfnamefont {W.~S.}\ \bibnamefont
  {Kolthammer}}, \ and\ \bibinfo {author} {\bibfnamefont {I.~A.}\ \bibnamefont
  {Walmsley}},\ }\href@noop {} {\bibfield  {journal} {\bibinfo  {journal}
  {arXiv.org}\ } (\bibinfo {year} {2016})},\ \Eprint
  {http://arxiv.org/abs/1603.08788} {1603.08788} \BibitemShut {NoStop}%
\bibitem [{\citenamefont {Bennett}\ \emph {et~al.}(1993)\citenamefont
  {Bennett}, \citenamefont {Brassard}, \citenamefont {Crepeau}, \citenamefont
  {Jozsa}, \citenamefont {Peres},\ and\ \citenamefont {Wootters}}]{bennett93}%
  \BibitemOpen
  \bibfield  {author} {\bibinfo {author} {\bibfnamefont {C.~H.}\ \bibnamefont
  {Bennett}}, \bibinfo {author} {\bibfnamefont {G.}~\bibnamefont {Brassard}},
  \bibinfo {author} {\bibfnamefont {C.}~\bibnamefont {Crepeau}}, \bibinfo
  {author} {\bibfnamefont {R.}~\bibnamefont {Jozsa}}, \bibinfo {author}
  {\bibfnamefont {A.}~\bibnamefont {Peres}}, \ and\ \bibinfo {author}
  {\bibfnamefont {W.~K.}\ \bibnamefont {Wootters}},\ }\href {\doibase
  10.1103/PhysRevLett.70.1895} {\bibfield  {journal} {\bibinfo  {journal}
  {Phys. Rev. Lett.}\ }\textbf {\bibinfo {volume} {70}},\ \bibinfo {pages}
  {1895} (\bibinfo {year} {1993})}\BibitemShut {NoStop}%
\bibitem [{\citenamefont {Bouwmeester}\ \emph {et~al.}(1997)\citenamefont
  {Bouwmeester}, \citenamefont {Pan}, \citenamefont {Mattle}, \citenamefont
  {Eibl}, \citenamefont {Weinfurter},\ and\ \citenamefont
  {Zeilinger}}]{bouwmeester97}%
  \BibitemOpen
  \bibfield  {author} {\bibinfo {author} {\bibfnamefont {D.}~\bibnamefont
  {Bouwmeester}}, \bibinfo {author} {\bibfnamefont {J.}~\bibnamefont {Pan}},
  \bibinfo {author} {\bibfnamefont {K.}~\bibnamefont {Mattle}}, \bibinfo
  {author} {\bibfnamefont {M.}~\bibnamefont {Eibl}}, \bibinfo {author}
  {\bibfnamefont {H.}~\bibnamefont {Weinfurter}}, \ and\ \bibinfo {author}
  {\bibfnamefont {A.}~\bibnamefont {Zeilinger}},\ }\href {\doibase
  10.1038/37539} {\bibfield  {journal} {\bibinfo  {journal} {Nature}\ }\textbf
  {\bibinfo {volume} {390}},\ \bibinfo {pages} {575} (\bibinfo {year}
  {1997})}\BibitemShut {NoStop}%
\bibitem [{\citenamefont {Kok}\ and\ \citenamefont {Braunstein}(2000)}]{kok00}%
  \BibitemOpen
  \bibfield  {author} {\bibinfo {author} {\bibfnamefont {P.}~\bibnamefont
  {Kok}}\ and\ \bibinfo {author} {\bibfnamefont {S.~L.}\ \bibnamefont
  {Braunstein}},\ }\href {\doibase 10.1103/PhysRevA.61.042304} {\bibfield
  {journal} {\bibinfo  {journal} {Phys. Rev. A}\ }\textbf {\bibinfo {volume}
  {61}},\ \bibinfo {pages} {042304} (\bibinfo {year} {2000})}\BibitemShut
  {NoStop}%
\bibitem [{\citenamefont {Browne}\ and\ \citenamefont
  {Rudolph}(2005)}]{Browne05}%
  \BibitemOpen
  \bibfield  {author} {\bibinfo {author} {\bibfnamefont {D.~E.}\ \bibnamefont
  {Browne}}\ and\ \bibinfo {author} {\bibfnamefont {T.}~\bibnamefont
  {Rudolph}},\ }\href {\doibase 10.1103/PhysRevLett.95.010501} {\bibfield
  {journal} {\bibinfo  {journal} {Phys. Rev. Lett.}\ }\textbf {\bibinfo
  {volume} {95}},\ \bibinfo {pages} {010501} (\bibinfo {year}
  {2005})}\BibitemShut {NoStop}%
\bibitem [{\citenamefont {Briegel}\ \emph {et~al.}(1998)\citenamefont
  {Briegel}, \citenamefont {D{\"u}r}, \citenamefont {Cirac},\ and\
  \citenamefont {Zoller}}]{briegel98}%
  \BibitemOpen
  \bibfield  {author} {\bibinfo {author} {\bibfnamefont {H.~J.}\ \bibnamefont
  {Briegel}}, \bibinfo {author} {\bibfnamefont {W.}~\bibnamefont {D{\"u}r}},
  \bibinfo {author} {\bibfnamefont {J.~I.}\ \bibnamefont {Cirac}}, \ and\
  \bibinfo {author} {\bibfnamefont {P.}~\bibnamefont {Zoller}},\ }\href
  {\doibase 10.1103/PhysRevLett.81.5932} {\bibfield  {journal} {\bibinfo
  {journal} {Phys. Rev. Lett.}\ }\textbf {\bibinfo {volume} {81}},\ \bibinfo
  {pages} {5932} (\bibinfo {year} {1998})}\BibitemShut {NoStop}%
\bibitem [{\citenamefont {D{\"u}r}\ \emph {et~al.}(1999)\citenamefont
  {D{\"u}r}, \citenamefont {Briegel}, \citenamefont {Cirac},\ and\
  \citenamefont {Zoller}}]{Dur99}%
  \BibitemOpen
  \bibfield  {author} {\bibinfo {author} {\bibfnamefont {W.}~\bibnamefont
  {D{\"u}r}}, \bibinfo {author} {\bibfnamefont {H.~J.}\ \bibnamefont
  {Briegel}}, \bibinfo {author} {\bibfnamefont {J.~I.}\ \bibnamefont {Cirac}},
  \ and\ \bibinfo {author} {\bibfnamefont {P.}~\bibnamefont {Zoller}},\ }\href
  {\doibase 10.1103/PhysRevA.59.169} {\bibfield  {journal} {\bibinfo  {journal}
  {Phys. Rev. A}\ }\textbf {\bibinfo {volume} {59}},\ \bibinfo {pages} {169}
  (\bibinfo {year} {1999})}\BibitemShut {NoStop}%
\bibitem [{\citenamefont {Kok}\ \emph {et~al.}(2003)\citenamefont {Kok},
  \citenamefont {Williams},\ and\ \citenamefont {Dowling}}]{Kok03b}%
  \BibitemOpen
  \bibfield  {author} {\bibinfo {author} {\bibfnamefont {P.}~\bibnamefont
  {Kok}}, \bibinfo {author} {\bibfnamefont {C.~P.}\ \bibnamefont {Williams}}, \
  and\ \bibinfo {author} {\bibfnamefont {J.~P.}\ \bibnamefont {Dowling}},\
  }\href {\doibase 10.1103/PhysRevA.68.022301} {\bibfield  {journal} {\bibinfo
  {journal} {Phys. Rev. A}\ }\textbf {\bibinfo {volume} {68}},\ \bibinfo
  {pages} {022301} (\bibinfo {year} {2003})}\BibitemShut {NoStop}%
\bibitem [{\citenamefont {Weinfurter}(1994)}]{Weinfurter94}%
  \BibitemOpen
  \bibfield  {author} {\bibinfo {author} {\bibfnamefont {H.}~\bibnamefont
  {Weinfurter}},\ }\href {\doibase 10.1238/Physica.Topical.076a00203}
  {\bibfield  {journal} {\bibinfo  {journal} {Eur. Phys. Lett.}\ }\textbf
  {\bibinfo {volume} {25}},\ \bibinfo {pages} {559} (\bibinfo {year}
  {1994})}\BibitemShut {NoStop}%
\bibitem [{\citenamefont {Braunstein}\ and\ \citenamefont
  {Mann}(1995)}]{Braunstein95b}%
  \BibitemOpen
  \bibfield  {author} {\bibinfo {author} {\bibfnamefont {S.~L.}\ \bibnamefont
  {Braunstein}}\ and\ \bibinfo {author} {\bibfnamefont {A.}~\bibnamefont
  {Mann}},\ }\href {\doibase 10.1103/PhysRevA.51.R1727} {\bibfield  {journal}
  {\bibinfo  {journal} {Phys. Rev. A}\ }\textbf {\bibinfo {volume} {51}},\
  \bibinfo {pages} {R1727} (\bibinfo {year} {1995})}\BibitemShut {NoStop}%
\bibitem [{\citenamefont {Vaidman}\ and\ \citenamefont
  {Yoran}(1999)}]{Vaidman99}%
  \BibitemOpen
  \bibfield  {author} {\bibinfo {author} {\bibfnamefont {L.}~\bibnamefont
  {Vaidman}}\ and\ \bibinfo {author} {\bibfnamefont {N.}~\bibnamefont
  {Yoran}},\ }\href {\doibase 10.1103/PhysRevA.59.116} {\bibfield  {journal}
  {\bibinfo  {journal} {Phys. Rev. A}\ }\textbf {\bibinfo {volume} {59}},\
  \bibinfo {pages} {116} (\bibinfo {year} {1999})}\BibitemShut {NoStop}%
\bibitem [{\citenamefont {L{\"u}tkenhaus}\ \emph {et~al.}(1999)\citenamefont
  {L{\"u}tkenhaus}, \citenamefont {Calsamiglia},\ and\ \citenamefont
  {Suominen}}]{Lutkenhaus99}%
  \BibitemOpen
  \bibfield  {author} {\bibinfo {author} {\bibfnamefont {N.}~\bibnamefont
  {L{\"u}tkenhaus}}, \bibinfo {author} {\bibfnamefont {J.}~\bibnamefont
  {Calsamiglia}}, \ and\ \bibinfo {author} {\bibfnamefont {K.~A.}\ \bibnamefont
  {Suominen}},\ }\href {\doibase 10.1103/PhysRevA.59.3295} {\bibfield
  {journal} {\bibinfo  {journal} {Phys. Rev. A}\ }\textbf {\bibinfo {volume}
  {59}},\ \bibinfo {pages} {3295} (\bibinfo {year} {1999})}\BibitemShut
  {NoStop}%
\bibitem [{\citenamefont {Helstrom}(1973)}]{Helstrom73}%
  \BibitemOpen
  \bibfield  {author} {\bibinfo {author} {\bibfnamefont {C.~W.}\ \bibnamefont
  {Helstrom}},\ }\href {\doibase 10.1007/BF00687093} {\bibfield  {journal}
  {\bibinfo  {journal} {Int. J. Theor. Phys.}\ }\textbf {\bibinfo {volume}
  {8}},\ \bibinfo {pages} {361} (\bibinfo {year} {1973})}\BibitemShut {NoStop}%
\bibitem [{\citenamefont {Ragy}\ \emph {et~al.}(2016)\citenamefont {Ragy},
  \citenamefont {Jarzyna},\ and\ \citenamefont
  {Demkowicz-Dobrza{\'{n}}ski}}]{Ragy16}%
  \BibitemOpen
  \bibfield  {author} {\bibinfo {author} {\bibfnamefont {S.}~\bibnamefont
  {Ragy}}, \bibinfo {author} {\bibfnamefont {M.}~\bibnamefont {Jarzyna}}, \
  and\ \bibinfo {author} {\bibfnamefont {R.}~\bibnamefont
  {Demkowicz-Dobrza{\'{n}}ski}},\ }\href {\doibase 10.1103/PhysRevA.94.052108}
  {\bibfield  {journal} {\bibinfo  {journal} {Phys. Rev. A}\ }\textbf {\bibinfo
  {volume} {94}},\ \bibinfo {pages} {052108} (\bibinfo {year}
  {2016})}\BibitemShut {NoStop}%
\bibitem [{\citenamefont {Holevo}(1982)}]{Holevo82}%
  \BibitemOpen
  \bibfield  {author} {\bibinfo {author} {\bibfnamefont {A.~S.}\ \bibnamefont
  {Holevo}},\ }\href@noop {} {\emph {\bibinfo {title} {{Probabilistic and
  Statistical Aspects of Quantum Theory}}}}\ (\bibinfo  {publisher} {North
  Holland, Amsterdam},\ \bibinfo {year} {1982})\BibitemShut {NoStop}%
\bibitem [{\citenamefont {Kitaev}(1997)}]{Kitaev97}%
  \BibitemOpen
  \bibfield  {author} {\bibinfo {author} {\bibfnamefont {A.~Y.}\ \bibnamefont
  {Kitaev}},\ }\href@noop {} {\bibfield  {journal} {\bibinfo  {journal} {Russ.
  Math. Surv.}\ }\textbf {\bibinfo {volume} {52}},\ \bibinfo {pages} {1191}
  (\bibinfo {year} {1997})}\BibitemShut {NoStop}%
\bibitem [{\citenamefont {Wang}\ \emph {et~al.}(2013)\citenamefont {Wang},
  \citenamefont {Berry}, \citenamefont {de~Oliveira},\ and\ \citenamefont
  {Sanders}}]{Wang13}%
  \BibitemOpen
  \bibfield  {author} {\bibinfo {author} {\bibfnamefont {D.-S.}\ \bibnamefont
  {Wang}}, \bibinfo {author} {\bibfnamefont {D.~W.}\ \bibnamefont {Berry}},
  \bibinfo {author} {\bibfnamefont {M.~C.}\ \bibnamefont {de~Oliveira}}, \ and\
  \bibinfo {author} {\bibfnamefont {B.~C.}\ \bibnamefont {Sanders}},\ }\href
  {\doibase 10.1103/PhysRevLett.111.130504} {\bibfield  {journal} {\bibinfo
  {journal} {Physical Review Letters}\ }\textbf {\bibinfo {volume} {111}},\
  \bibinfo {pages} {81} (\bibinfo {year} {2013})}\BibitemShut {NoStop}%
\bibitem [{\citenamefont {Wang}\ and\ \citenamefont {Sanders}(2018)}]{Wang18}%
  \BibitemOpen
  \bibfield  {author} {\bibinfo {author} {\bibfnamefont {D.-S.}\ \bibnamefont
  {Wang}}\ and\ \bibinfo {author} {\bibfnamefont {B.~C.}\ \bibnamefont
  {Sanders}},\ }\href {\doibase 10.1088/1367-2630/17/4/043004} {\bibfield
  {journal} {\bibinfo  {journal} {New Journal of Physics}\ }\textbf {\bibinfo
  {volume} {17}},\ \bibinfo {pages} {1} (\bibinfo {year} {2018})}\BibitemShut
  {NoStop}%
\bibitem [{\citenamefont {Campbell}(2017)}]{Campbell17}%
  \BibitemOpen
  \bibfield  {author} {\bibinfo {author} {\bibfnamefont {E.}~\bibnamefont
  {Campbell}},\ }\href {\doibase 10.1103/PhysRevA.95.042306} {\bibfield
  {journal} {\bibinfo  {journal} {Physical Review A}\ }\textbf {\bibinfo
  {volume} {95}},\ \bibinfo {pages} {81} (\bibinfo {year} {2017})}\BibitemShut
  {NoStop}%
\bibitem [{\citenamefont {Campbell}\ and\ \citenamefont
  {O{\textquoteright}Gorman}(2018)}]{Campbell18}%
  \BibitemOpen
  \bibfield  {author} {\bibinfo {author} {\bibfnamefont {E.~T.}\ \bibnamefont
  {Campbell}}\ and\ \bibinfo {author} {\bibfnamefont {J.}~\bibnamefont
  {O{\textquoteright}Gorman}},\ }\href {\doibase 10.1088/2058-9565/1/1/015007}
  {\bibfield  {journal} {\bibinfo  {journal} {Quantum Sci Technol}\ }\textbf
  {\bibinfo {volume} {1}},\ \bibinfo {pages} {1} (\bibinfo {year}
  {2018})}\BibitemShut {NoStop}%
\bibitem [{\citenamefont {Wootters}(1981)}]{Wootters81}%
  \BibitemOpen
  \bibfield  {author} {\bibinfo {author} {\bibfnamefont {W.~K.}\ \bibnamefont
  {Wootters}},\ }\href {\doibase 10.1103/PhysRevD.23.357} {\bibfield  {journal}
  {\bibinfo  {journal} {Physical Review D}\ }\textbf {\bibinfo {volume} {23}},\
  \bibinfo {pages} {357} (\bibinfo {year} {1981})}\BibitemShut {NoStop}%
\end{thebibliography}

%

\end{document}